\documentclass[aps,prd,twocolumn,superscriptaddress,a4paper]{revtex4}

\usepackage{graphicx}
\usepackage{eurosym}

\begin{document}


\title{The SIMPLE Phase II Dark Matter Search}


\author{M. Felizardo}
\affiliation{Department of Physics,
Universidade Nova de Lisboa, 2829-516 Caparica, Portugal} \affiliation{Centro de F\'isica
Nuclear, Universidade de Lisboa, 1649--003 Lisbon, Portugal} \affiliation{Instituto Tecnol\'ogico e Nuclear, IST, Universidade T\'ecnica de Lisboa, EN 10, 2686-953 Sacav\'em, Portugal}
\author{TA Girard}
\email[]{correspondent: criodets@cii.fc.ul.pt}
\affiliation{Centro de F\'isica Nuclear, Universidade de Lisboa, 1649--003 Lisbon, Portugal} \affiliation{Department of Physics, Universidade Lisboa, 1749-016 Lisbon, Portugal}
\author{T. Morlat}
\affiliation{Ecole Normale Superieur de Montrouge, 1 Rue Aurice Arnoux, 92120 Montrouge, France}
\author{A.C. Fernandes} \affiliation{Instituto Tecnol\'ogico e Nuclear, IST, Universidade T\'ecnica de Lisboa, EN 10, 2686-953 Sacav\'em, Portugal} \affiliation{Centro de F\'isica Nuclear, Universidade de
Lisboa, 1649--003 Lisbon, Portugal}
\author{A.R. Ramos} \affiliation{Instituto Tecnol\'ogico e Nuclear, IST, Universidade T\'ecnica de Lisboa, EN 10, 2686-953 Sacav\'em, Portugal} \affiliation{Centro de F\'isica Nuclear, Universidade de Lisboa, 1649--003 Lisbon, Portugal}
\author{J.G. Marques}
\affiliation{Instituto Tecnol\'ogico e Nuclear, IST, Universidade T\'ecnica de Lisboa, EN 10, 2686-953 Sacav\'em, Portugal} \affiliation{Centro de F\'isica Nuclear, Universidade de Lisboa, 1649--003 Lisbon, Portugal}
\author{A. Kling}
\affiliation{Instituto Tecnol\'ogico e Nuclear, IST, Universidade T\'ecnica de Lisboa, EN 10, 2686-953 Sacav\'em, Portugal} \affiliation{Centro de F\'isica Nuclear, Universidade de Lisboa, 1649--003 Lisbon, Portugal}
\author{J. Puibasset }
\affiliation{Centre de Recherche sur la Mati\`ere Divis\'ee - CNRS, Universit\'e de Orl\'eans, 45071 Orl\'eans, France}
\author{M. Auguste}
\affiliation{Laboratoire Souterrain \`a Bas Bruit (UMS 3538 UNS/UAPV/CNRS), 84400 Rustrel--Pays d'Apt, France}
\author{D. Boyer}
\affiliation{Laboratoire Souterrain \`a Bas Bruit (UMS 3538 UNS/UAPV/CNRS), 84400 Rustrel--Pays d'Apt, France}
\author{A. Cavaillou}
\affiliation{Laboratoire Souterrain \`a Bas Bruit (UMS 3538 UNS/UAPV/CNRS), 84400 Rustrel--Pays d'Apt, France}
\author{J. Poupeney}
\affiliation{Laboratoire Souterrain \`a Bas Bruit (UMS 3538 UNS/UAPV/CNRS), 84400 Rustrel--Pays d'Apt, France}
\author{C. Sudre}
\affiliation{Laboratoire Souterrain \`a Bas Bruit (UMS 3538 UNS/UAPV/CNRS), 84400 Rustrel--Pays d'Apt, France}
\author{F.P. Carvalho}
\affiliation{Instituto Tecnol\'ogico e Nuclear, IST, Universidade T\'ecnica de Lisboa, EN 10, 2686-953 Sacav\'em, Portugal}
\author{M.I. Prud\^encio}
\affiliation{Instituto Tecnol\'ogico e Nuclear, IST, Universidade T\'ecnica de Lisboa, EN 10, 2686-953 Sacav\'em, Portugal}
\author{R. Marques}
\affiliation{Instituto Tecnol\'ogico e Nuclear, IST, Universidade T\'ecnica de Lisboa, EN 10, 2686-953 Sacav\'em, Portugal}

\collaboration{The SIMPLE collaboration} \noaffiliation

\date{\today}

\begin{abstract}
Phase II of SIMPLE (Superheated Instrument for Massive ParticLe Experiments) searched for astroparticle dark matter using superheated liquid C$_{2}$ClF$_{5}$ droplet detectors. Each droplet generally requires an energy deposition with linear energy transfer (LET) $\gtrsim$ 150 keV/$\mu$m for a liquid-to-gas phase transition, providing an intrinsic rejection against minimum ionizing particles of order 10$^{-10}$, and reducing the backgrounds to primarily $\alpha$ and neutron-induced recoil events. The droplet phase transition generates a millimetric-sized gas bubble which is recorded by acoustic means. We describe the SIMPLE detectors, their acoustic instrumentation, and the characterizations, signal analysis and data selection which yield a particle-induced, "true nucleation" event detection efficiency of better than 97\% at a 95\% C.L. The recoil-$\alpha$ event discrimination, determined using detectors first irradiated with neutrons and then doped with alpha emitters, provides a recoil identification of better than 99\%; it differs from those of COUPP and PICASSO primarily as a result of their different liquids with lower critical LETs. The science measurements, comprising two shielded arrays of fifteen detectors each and a total exposure of 27.77 kgd, are detailed. Removal of the 1.94 kgd Stage 1 installation period data, which had previously been mistakenly included in the data, reduces the science exposure from 20.18 to 18.24 kgd and provides new contour minima of $\sigma_{p}$ = 4.3 $\times$ 10$^{-3}$ pb at 35 GeV/c$^{2}$ in the spin-dependent sector of WIMP-proton interactions and $\sigma_{N}$ = 3.6 $\times$ 10$^{-6}$ pb at 35 GeV/c$^{2}$ in the spin-independent sector. These results are examined with respect to the fluorine spin and halo parameters used in the previous data analysis.
\end{abstract}

\pacs{}

\maketitle


\section{INTRODUCTION}

SIMPLE (Superheated Instrument for Massive ParticLe Experiments) is one of only three experiments \cite{prl2,coupp2012,pic2012} world-wide in search of evidence of astroparticle dark matter (WIMPs) using freon-loaded superheated liquid (SHL) detectors. A SHL detector consists of either droplet dispersion (SDD) or bulk superheated liquid bubble chambers, which may undergo a transition to the gas phase upon energy deposition by incident radiation depending on two criteria \cite{seitz}: (i) the energy deposited must be greater than a thermodynamic minimum, and (ii) this energy must be deposited within a thermodynamically-defined maximum distance within the liquid.

SIMPLE employs chlorofluorocarbon C$_{2}$ClF$_{5}$, for which the two conditions together generally require at standard operating pressures and temperature a linear energy transfer (LET) $\gtrsim$ 150 keV/$\mu$m for a bubble nucleation. This renders the detector effectively insensitive to the majority of traditional detector backgrounds which complicate more conventional dark matter search detectors (including $\beta$'s, $\gamma$'s below 6 MeV, and cosmic muons). This intrinsic insensitivity is significant, comprising a rejection factor superior to that of other search techniques by 1-5 orders of magnitude. All three SHL projects have demonstrated a potential to achieve competitive results with relatively small measurement exposures (kg of detector active mass $\times$ days of measurement time).

In 2012, SIMPLE reported the final results of its Phase II measurements \cite{prl2} using SDDs with a total 27.77 kgd exposure (of which 20.18 kgd were taken to be science), conducted in the 60 m$^{3}$ GESA site (505 m depth) of the Laboratoire Souterrain \`a Bas Bruit (LSBB) in southern France \cite{lsbb}. With an 8 keV recoil energy threshold imposed by temperature and pressure control, the upper limit in the spin-independent sector impacted on current areas identified by several search experiments as containing a possible WIMP presence, while providing in the spin-dependent sector one of the most restrictive limits against a WIMP-proton coupling at the time.

Subsequent to its initial report \cite{prl1}, SIMPLE received various criticisms based on either its detector longevity \cite{cmt1} or signal analysis \cite{cmt2}, which were publicly addressed in two replies \cite{reply1,reply2}. Nevertheless, the results continue to receive comments as being "controversial" \cite{coupp2012}; we here elaborate in greater detail on the Phase II program, which was not amiable to the constraints of Letter publications, towards a further clarification of various issues. We also remove from the previously reported results Stage 1 installation data (taken with the neutron shield partially dismantled, intended for calibrations) which was recently discovered to have been inadvertently included with science. This correction reduces the net science exposure to 18.24 kgd and yields slightly improved constraints in both WIMP sectors.

The basic detector fabrication and instrumentation are described in Sec. II. Section III describes the basic sensitivities and signal analysis protocol, together with the calibrations which provide the basis for identification of recoil events, definition of the recoil acceptance window, and recoil-$\alpha$ discrimination. The Phase II experimental setup and detector installation is detailed in Section IV; the measurement data acquisition and its analysis is described in Sec. V, to include its background estimates and correction for the inadvertent inclusion of the Stage 1 installation data. Section VI reviews the interpretation of the results, and re-examines their variations in light of more accurate fluorine spin calculations and variations in the standard halo model parameters. A summary is provided in Sec. VII.

\section{THE DETECTOR CONSTRUCT}

\subsection{Fabrication}

\subsubsection{Gel}

SIMPLE SDD fabrications generally proceed on the basis of density-matching the C$_{2}$ClF$_{5}$ liquid with a 1.3 g/cm$^{3}$ food-based gel with low U/Th contamination: a significant difference in gel and liquid densities results in inhomogeneous distributions of differential droplet sizes within the detector. All SDDs were created according to a "standard fabrication" protocol developed for device uniformity, sensitivity, response and longevity during extensive experimental device R\&D during both Phase I \& II \cite{jptese,njp,tmtese,tomonim}. The "standard fabrication" gel composition is 1.71\% gelatin, 4.18\% polyvinylpyrrodine (PVP), 15.48\% bi-distilled water and 78.16\% glycerin: all ingredients are biologically-clean food products.

The protocol begins by combining powdered gelatin (Sigma Aldrich G-1890 Type A), bi-distilled water and pre-eluted ion exchange resins for actinide removal, which is left at 45$^{\circ}$C with slow agitation for 12-15 hr to homogenize the solution. Separately, PVP (Sigma Aldrich PVP-40T) and exchange resins are added to bi-distilled water, and agitated at ~65$^{\circ}$C for 12-15 hr. Resins and glycerin (Riedel-de-Haën Nº 33224) are combined separately, and left with medium agitation at $\sim$ 50$^{\circ}$C for 12-15 hr.

The PVP solution is then slowly added to the gel solution ("concentrated gel"), and slowly agitated at 55-60$^{o}$C for 2 hr. The resins are next separately removed from both the concentrated gel and glycerin by filtering (Whatman, Ref: 6725-5002A), and the two combined at $\sim$ 60$^{o}$C: the mix is outgassed at $\sim$ 70$^{o}$C, and then foam aspirated to eliminate trapped air bubbles. The solution is left at 48$^{o}$C for 14 hr with slow agitation to prevent bubble formation.

\subsubsection{Droplet Suspension}

Following transfer of the gel to the detector bottle, each bottle is first weighed (mg precision) and then installed within a glass beaker surrounded by a refrigerated copper serpentine positioned on a hotplate within a hyperbaric chamber.

\begin{figure}[h]
  \includegraphics[width=8 cm]{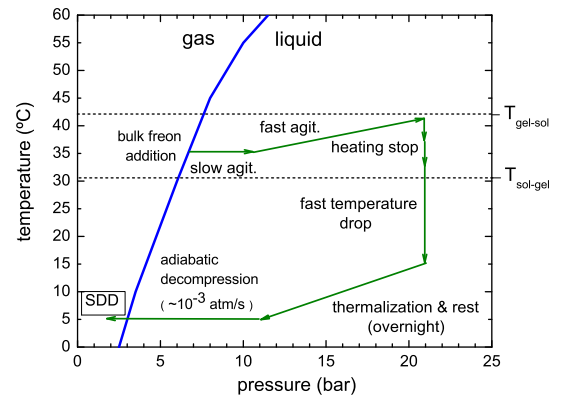}\\
  \caption{phase diagram of the fabrication protocol for a C$_{2}$ClF$_{5}$ SDD. The indicated sol-gel transition denotes the onset of a gel-like di-phasic system containing both liquid and solid phases whose morphology is a continuous polymer network.}
  \label{fig1}
\end{figure}

The freon injection protocol is shown in Fig. 1. Once the temperature is stabilized at 35$^{o}$C, the pressure is quickly raised to just above the vapor pressure ($\sim$ 11 bar) of the C$_{2}$ClF$_{5}$ with continued slow agitation. After thermalization, the agitation is stopped and the liquid C$_{2}$ClF$_{5}$ injected into the gel through a flowline immersed in ice to condense and distill it at the same time, which contains a 0.2 $\mu$m microsyringe filter (Gelman Acrodisc CR PTFE 4552T) to remove impurities.

\begin{figure}[h]
  \includegraphics[width=6 cm]{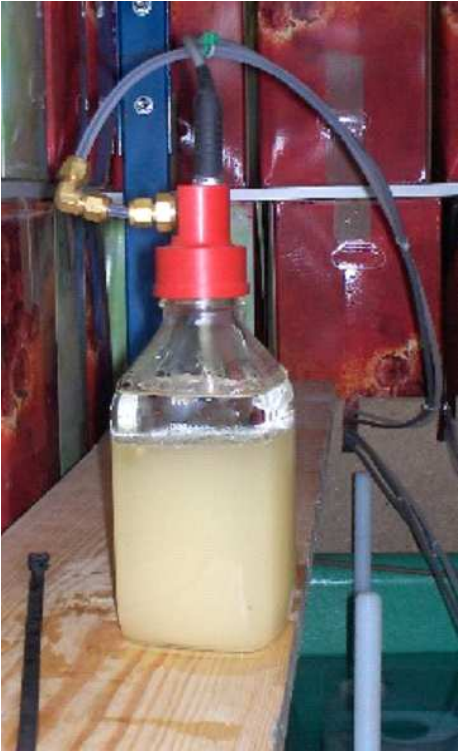}\\
  \caption{a science C$_{2}$ClF$_{5}$ SDD during installation; the bottle is a squared 10 $\times$ 10 cm$^{2}$ base, 12 cm tall Schott glass containing $\sim$ 900 ml of freon+gel covered by a 6 cm glycerin layer. The cap contains the microphone and feedthroughs for the pressure (horizontal) and microphone (vertical) electronics. }
  \label{fig1}
\end{figure}

Once injected, the pressure is quickly raised to 21 bar to prevent the droplets from rising to the gel surface, with a rapid agitation initiated to shear big droplets; simultaneously, the temperature is raised to 39$^{o}$C to create a temperature gradient inside the gel matrix and to permit dispersion of the droplets. After 15 minutes, the temperature is reduced to 37$^{o}$C for 30 min, then reduced to 35$^{o}$C for 4 hr with pressure and agitation unchanged to continue fractionating the liquid into smaller droplets. Finally, the heating is stopped: the temperature decreases until the sol-gel transition is crossed, during which the agitation is maintained. Approximately 2 hr later, the droplet suspension is quickly cooled to 15$^{o}$C with the serpentine, and left to set for 40 minutes with decreasing agitation as the gel solidifies; once the agitation is stopped, the pressure is slowly reduced over 10 min to 11 bar, where it is maintained for $\sim$ 15 hours with the temperature set to the measurement run temperature. Finally, the chamber pressure is slowly reduced to atmospheric, and the detector removed, weighed to determine the active mass, topped with a 6 cm layer of glycerin, and transported to its measurement site where the instrumentation cap is installed, as shown in Fig. 2.

The distribution of droplet sizes in the SDD is adjustable, depending on the liquid and the fractionating time and speed in the fabrication process: longer fractionating times yield narrower distributions of smaller diameters; shorter times, broader distributions with larger diameter satellites. The presence of the PVP serves to reduce the surface tension of the liquid, facilitating the fractionating and eliminating satellites. Each SDD contains $\sim$ 10$^{7}$ droplets, depending on its liquid mass and fractionating time.

\subsection{Instrumentation}

New acoustic instrumentation replaced that of the previous Phase I, which was inherently incapable of signal discrimination (see Refs. \cite{njp, felizimprov,feliznew,mftese} for details). Each detector is hermetically sealed with a PVC cap containing feedthroughs for the instrumentation, a pressure transducer (Swagelok PTI-S-AG4-15-AQ) and a Panasonic (omnidirectional), high quality electret microphone cartridge (MCE-200) with a frequency range of 0.020-16 kHz (3 dB) and 0.01 Hz resolution, and a sensitivity of 7.9 mV/Pa at 1 kHz with 0.3 mV resolution. The microphone itself is submerged in a 6 cm thick glycerin layer covering the emulsion, encased in a latex sheath to prevent glycerin ingress.

Each microphone output is connected to a remotely-located, digitally-controlled, analog microphone preamplifier (Texas Instruments PGA2500), designed for use as a front end for high performance audio analog-to-digital converters, and characterized by low noise and harmonic distortion. The configurable output circuitry is not used since the voltage regulators de-stabilize the circuit and the protective diodes introduce noise into the system \cite{feliznew}. The preamplifier was initially employed with phantom power disconnected, but with both +5 V and -5 V power supplies; minor modifications of several capacitances and resistances in the recommended circuitry were introduced in the course of development \cite{feliznew}.

The preamp signals are stored in a pc-supported MATLAB platform in sequential files of variable duration via a National Instruments PCI-6251 I/O board, with the system resolving time defined by the selected sampling rate and number of detector channels in use.

\section{DETECTOR CHARACTERIZATIONS}

\subsection{Sensitivities}

The basic physics of the SDD operation is the same as bubble chambers, and described in detail in Ref. \cite{ghilea,derrico} and references therein. It is based on the "thermal spike" model of Seitz \cite{seitz} and can be divided into several stages \cite{sun,marty,plesset}. Initially, energy is deposited in a small volume of the liquid, producing a localized, high temperature region (the "thermal spike"), the sudden expansion of which produces a shock wave in the surrounding liquid. In this stage, the temperature and pressure of the liquid within the shock enclosure exceed the critical temperature and pressures: there is no distinction between liquid and vapor, and no bubble.

As the energy is transmitted from the thermalized region to the surrounding medium through shock propagation and heat conduction, the temperature and pressure of the fluid within the shock enclosure decrease, the expansion process slows and the shock wave decays. As the temperature and pressure reach the critical temperature and pressure, a vapor-liquid interface is formed which generates a proto-bubble. If the LET was sufficiently high, the vapor within the proto-bubble achieves a critical radius r$_{c}$ and a bubble forms via continued droplet evaporation continues; if the LET is insufficient, cavity growth is impeded by interfacial and viscous forces and conduction heat loss, and the proto-bubble collapses.

The thermodynamic conditions of the Seitz model for a proto-bubble to achieve r$_{c}$ require that, for a device operating pressure (P) and temperature (T), the deposited energy E$_{dep}$ must satisfy:

\begin{eqnarray}\label{vmin}
E_{dep} \geq E_{c} \\
dE_{dep}/dx \geq E_{c}/\Lambda r_{c} = LET_{c},
\end{eqnarray}

\noindent with

\begin{equation}\label{Ec}
E_{c} = 4\pi r_{c}^{2} (\sigma - T\frac{\partial \sigma}{\partial T}) + \frac{4}{3} \pi r_{c}^{3}\rho_{v}h_{lv} + \frac{4}{3}\pi r_{c}^{3}\Delta p ,
\end{equation}

\noindent where LET$_{c}$ is the critical LET for the bubble nucleation; r$_{c}$ = 2$\sigma$/$\Delta$p is the critical radius of proto-bubble formation, $\Delta$p = P$_{v}$-P with P$_{v}$ the saturated vapor pressure, $\sigma$ is the surface tension, $\rho_{v}$ is the saturated vapor density, and h$_{lv}$ is the latent heat of vaporization. Irreversible, generally smaller, terms (sound generation, conduction heat losses...) have been neglected. The $\Lambda$r$_{c}$ in Eq. (2) is the effective ionic energy deposition length, with $\Lambda$ = $\Lambda (T,P)$ the liquid-dependent nucleation parameter. Measurements in Phase I with $^{241}$Am $\alpha$-doped detector gels and temperature-ramping of the SDDs \cite{jptese} yielded $\Lambda$ = 1.40 $\pm$ 0.05 for C$_{2}$ClF$_{5}$, in agreement with an empirical $\Lambda$ = $\lambda$($\rho_{v}$/$\rho_{l}$)$^{1/3}$, with $\rho _{l}$ the liquid density and $\lambda$ = 4.3 \cite{harper} obtained experimentally. Together with thermodynamic parameters of C$_{2}$ClF$_{5}$ from Refs. \cite{duan,nist,bonin}, Eqs. (2)-(3) give LET$_{c}$ = 121 (176) keV/$\mu$m for C$_{2}$ClF$_{5}$ at 1 (2) bar and 9$^{o}$C.

Together, Eqs. (1)-(2) define the minimum energy threshold (E$_{thr}$) for bubble nucleation for each incident radiation. In general, the LET of $\beta$'s, cosmic-ray muons and $\gamma$'s of energy $<$ 6 MeV is well below LET$_{c}$ \cite{derrico, barnabe}, with a threshold sensitivity to these backgrounds occurring for a reduced superheat of s = [T-T$_{b}$]/[T$_{c}$-$T_{b}$] $\geq$ 0.5, where T$_{c}$, T$_{b}$ are the critical and boiling temperatures of the liquid, respectively \cite{derrico}. Of the customary backgrounds, only $\alpha$- and recoil-induced events are expected to contribute true bubble nucleation events to the data collection. The basic dependence on temperature and pressure of the energy threshold (E$_{thr}^{A}$) for bubble nucleation is shown in Fig. 3 for each liquid constituent of atomic mass A, calculated using Eqs. (1)-(3) with $\Lambda$ = 1.40, Refs. \cite{duan,nist,bonin}, and recoiling ion stopping powers calculated with SRIM \cite{srim}. Only energy depositions above the recoil curves can produce bubble nucleations.

\begin{figure}[h]
  \includegraphics[width=9 cm]{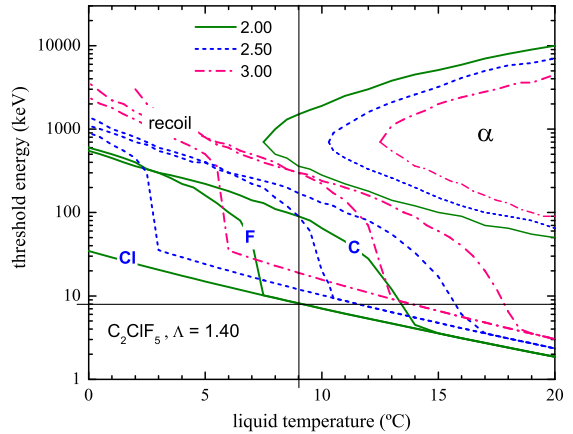}\\
  \caption{temperature variation of the threshold recoil energy with pressure for the three C$_{2}$ClF$_{5}$ constituents and $^{4}$He ions at 2.00 (solid), 2.50 (dotted) and 3.00 (dash-dot) bar, calculated with $\Lambda$ = 1.40. The vertical line denotes the standard operating temperature of the SDD; the horizontal, the 8 keV recoil threshold energy for 2 bar in the measurements. }
\label{fig3}
\end{figure}

The similarly calculated $\alpha$ response is also shown in Fig. 3, the "nose" of which corresponds to the $\alpha$ Bragg peak; note that SDDs at fixed temperature and pressure are sensitive to $\alpha$'s only within the "nose", which shifts to higher temperature with increased pressure (as also the recoil threshold energy curve). The $\alpha$ Bragg peak sets the temperature threshold for direct detection of $\alpha$'s; below this threshold, $\alpha$'s can only be detected through $\alpha$-induced nuclear recoils.

The metastability of a superheated liquid, as described by homogeneous nucleation theory, gives a stability limit of the liquid phase at approximately 90\% of the critical temperature for organic liquids at atmospheric pressure \cite{eber}: for moderately superheated liquids, the theoretical probability of spontaneous nucleation is negligible ($<$10$^{-1000}$ nucleations/kgd) and decreases with decreasing temperature. Given the purity and smooth droplet/gel interfaces of the SDDs, the inhomogeneous contribution is also negligible.

\subsection{Signal Analysis}

In addition to the $\alpha$ and recoil events, the acoustic data collection is also expected to contain a variety of background signal associated with the gel dynamics (fractures, trapped N$_{2}$ gas, ...) and various types of environmental noise.

The recorded bubble nucleation event signal is due to the acoustic pressure wave generated by the nucleation event, the amplitude $\mathcal{A}$ of which derives from the rate of the bubble expansion \cite{landau}:

\begin{equation}\label{K}
K = \mathcal{A}^{2}  = \frac{\rho_{l}}{4\pi c} \ddot{V}^{2} = \frac{4\pi\rho_{l}}{c} \frac{r^{6}}{\tau^{4}},
\end{equation}

\noindent where K is the acoustic power of the event, c is the speed of sound in the liquid, $\tau$ is the expansion time of the bubble, and V = $\frac{4\pi}{3}$r$^{3}$ is the droplet volume. An estimate of the pressure change at the microphone in a SDD is obtained \cite{plesset,felizfabvar}, again with Refs. \cite{duan,nist,bonin}, as $\mathcal{A}$ $\sim$ 6.2 $\times$ 10$^{2}$ $\mu$bar, so that the microphone sensitivity yields a nucleation signal $\mathcal{A} \sim$ 1000 mV at 1 kHz for droplet radii ~ 30 $\mu$m. The complete phase transition of a droplet results in a gas bubble harmonically oscillating about its equilibrium radius, with a resonant frequency depending on the elasticity of the gel as well as thermodynamic properties of both phases \cite{felizfabvar}, and is estimated at $\sim$ 700 Hz for C$_{2}$ClF$_{5}$ with typical operating parameters at T,P \cite{minnaert}.

Analysis of the signals generally follows a "standard protocol" developed in Ref. \cite{feliznew}, and derived from numerous calibrations with $\alpha$'s, neutrons and $\gamma$'s, as well as stimulated gel dynamics and environmental noises. The high concentration SDD response to neutrons has been extensively investigated using sources of Am/Be, $^{252}$Cf \cite{jptese,njp} and monochromatic low energy neutron beams \cite{itn}, and was studied during development of the device fabrication recipes in Phase I (see Refs. \cite{jptese,tmtese} for further details), as also Phase II with Am/Be irradiations. The $\alpha$ response has also been investigated by doping 250 ml "standard fabrication" devices with weak solutions of either $^{241}$Am or U$_{3}$0$_{8}$ both during and following fabrication.

In consequence of the calibration studies, a "true" nucleation event signal is characterized by a few ms time span, an amplitude of its PSD primary harmonic $\mathcal{A}$ $>$ 10 mV, a decay constant ($\tau_{0}$) of 5-40 ms, and frequency of its primary harmonic ($\mathcal{F}$) between 0.45-0.75 kHz \cite{feliznew}. Calibrations for a wide variety of acoustic backgrounds, of both gel and environmental origin, were similarly made; Table I contains examples of several. As seen, although the "true" bubble nucleation event $\tau_{0}$ span those of several background entries, the $\mathcal{F}$ for the most part differ.

The analysis protocol begins with an inspection of the signal records of each SDD response for raw signal rate and pressure evolution over the measurement period. An initial data set is formed by passing the data files through a pulse validation routine \cite{mftese} which tags signal events if their amplitudes exceed the noise level of the detector by 2 mV. Tagged signals in coincidence with a co-located freon-less device, which serves as an acoustic event monitor, are next rejected as also all candidate signals with less than five pulse spikes above threshold.

The signal waveforms of all surviving single events are next analyzed using characterization algorithms based on a Hilbert transform demodulation program to extract the $\mathcal{F}$ and $\tau_{0}$, which are used to order the priority of subsequent analysis.

\begin{table}[h]
\caption{\label{table} signal characteristics of bubble nucleation events, together with examples of some principle gel-associated and environmental noise backgrounds.}
\begin{ruledtabular}
\begin{tabular}{ccccccc}
  & $\tau_{0}$ (ms) & $\mathcal{F}$ (Hz)  \\
  \hline
   microleak & 2-60 & 2800-3500 \\
   fracture & 2-40 & 10-100 \\
   trapped N$_{2}$ release & 40-100  & 10-440 \\
   water bubbles & 2-25 & 1000-2000 \\
   cable vibrations & 10-20 & 750-1500 \\
   mechanical contacts & 5-40 & 100-450 \\
   local human activity & 24-33 & 750-1250 \\
   local vehicle movement & 17 & 2350 \\
   "true nucleation" event & 5-40 & 450-750 \\
\end{tabular}
\end{ruledtabular}
\end{table}

Finally, the Power Spectral Density (PSD) of each event falling within the $\mathcal{F}$-$\tau_{0}$ correlation window, as well as those at/near the borders, are inspected individually for comparison with the PSD of a true nucleation event template as shown in Fig. 4(a), which differs significantly from those of a variety of gel-associated acoustic backgrounds (such as trapped N$_{2}$ gas and gel fractures) which appear at lower $\mathcal{F}$ and $\tau_{0}$ \cite{feliznew}, and local acoustic backgrounds such as human activities and water bubbles (a preponderance of which were expected in the Phase II measurements as a result of the adjustment of the water circulation input of the temperature-controlling water pool to just above the pool water level in suppressing the atmospheric radon diffusion). An illustrative sample of an exhaustive "PSD gallery" of typical acoustic backgrounds associated with various gel dynamics and environmental noises is shown in Figs. 4(b)- 4(f). In most cases, over 100 events of each type were generated and unambiguously identified via event-by-event examination of the signal PSD; a binomial probability analysis gives a minimum efficiency ($\varepsilon$) for the identification of N particle-induced events in a sample of N with a confidence level C.L. during calibration measurements of $\varepsilon \geq$ (1-C.L.)$^{1/N+1}$ yielding $>$ 97\% at 95\% C.L.

\begin{figure*}
    \includegraphics[width=6 cm]{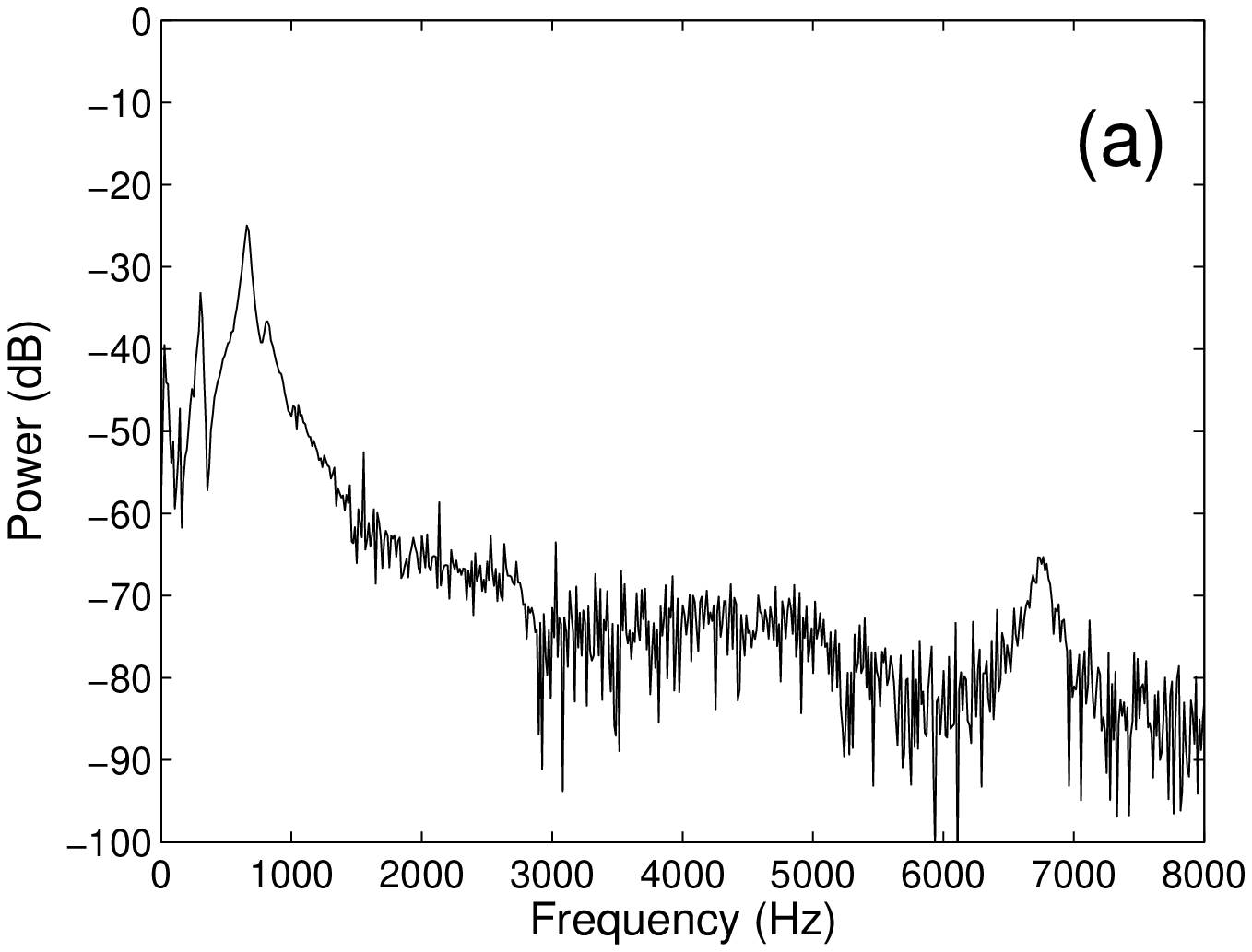}
    \includegraphics[width=6 cm]{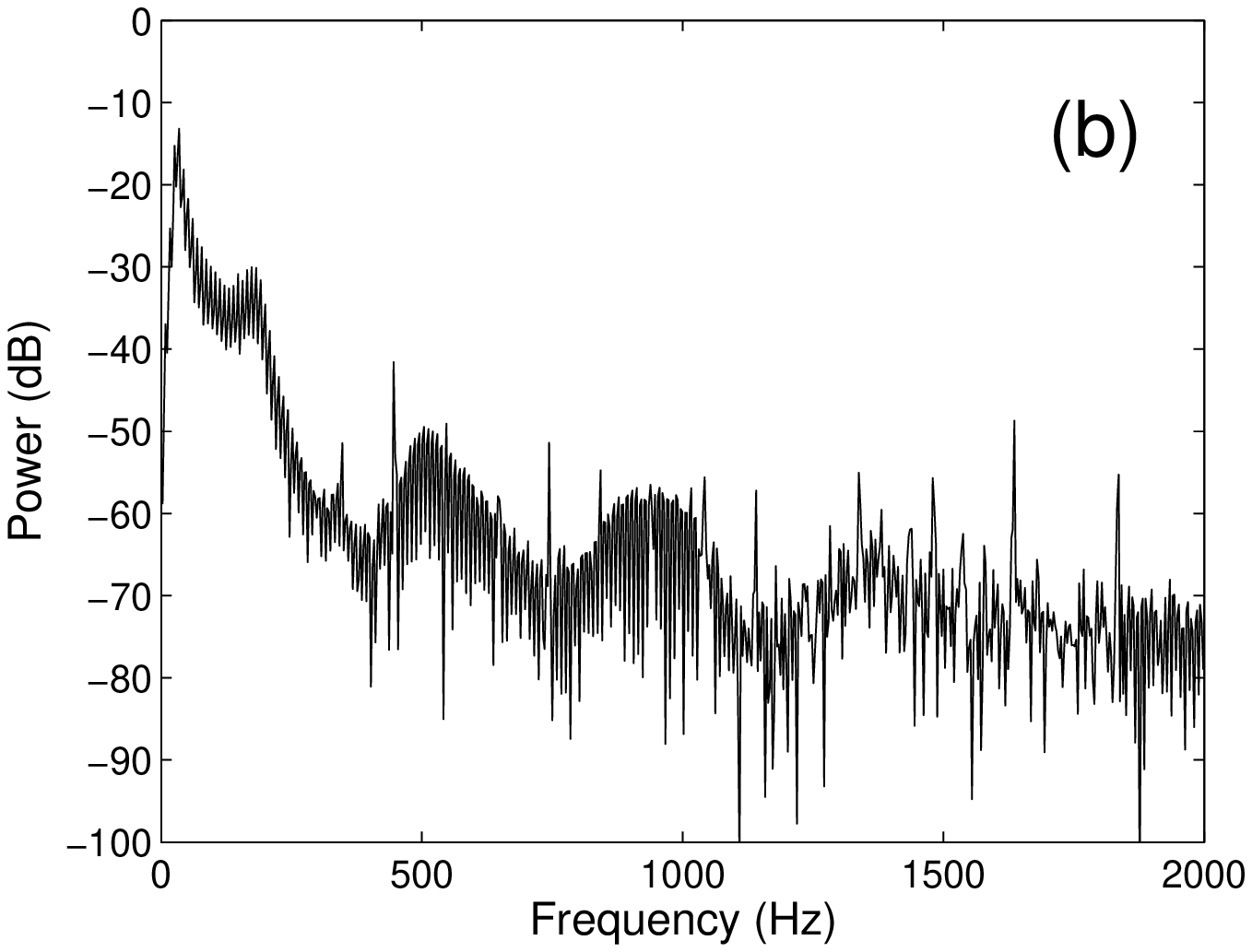}\\
    \includegraphics[width=6 cm]{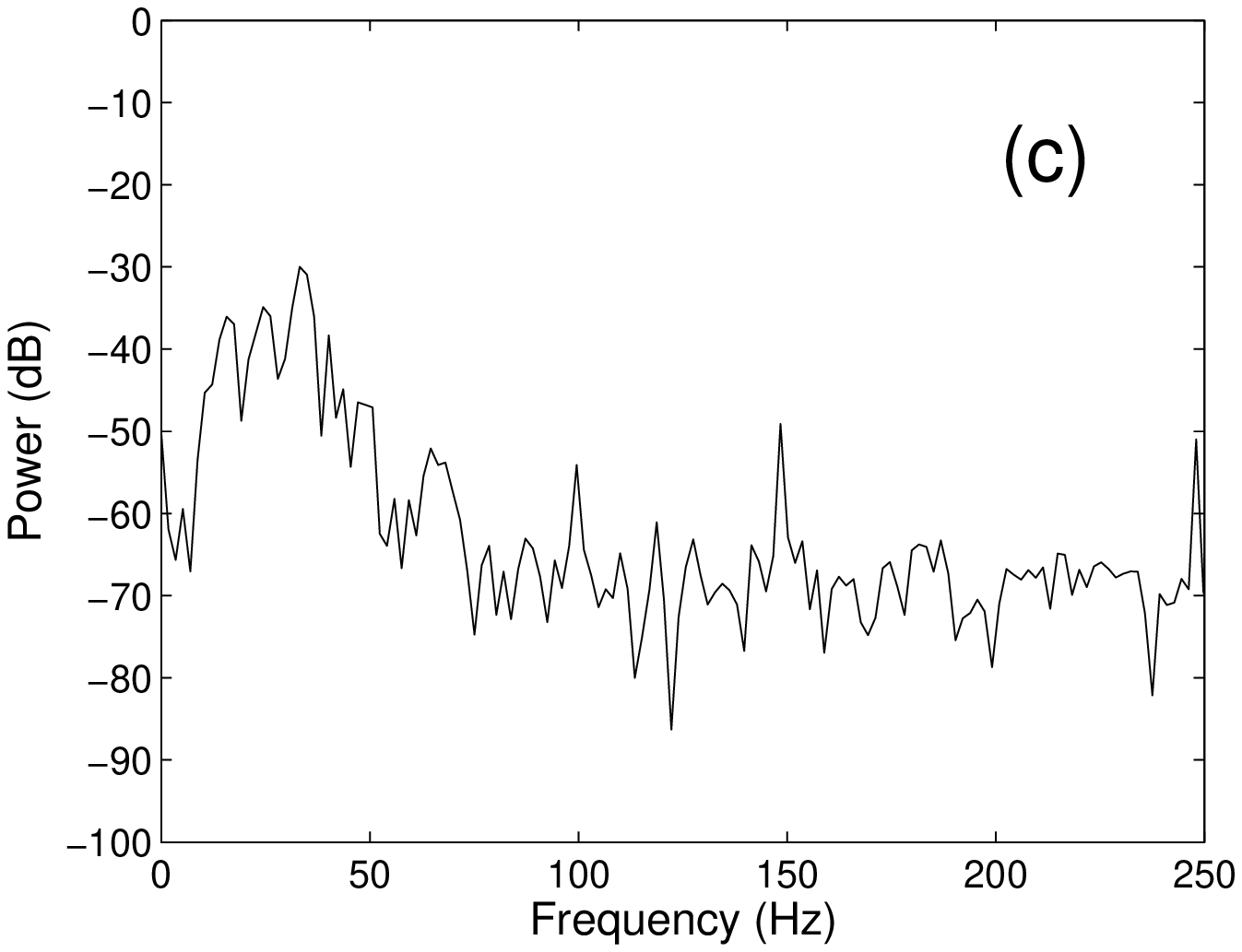}
    \includegraphics[width=6 cm]{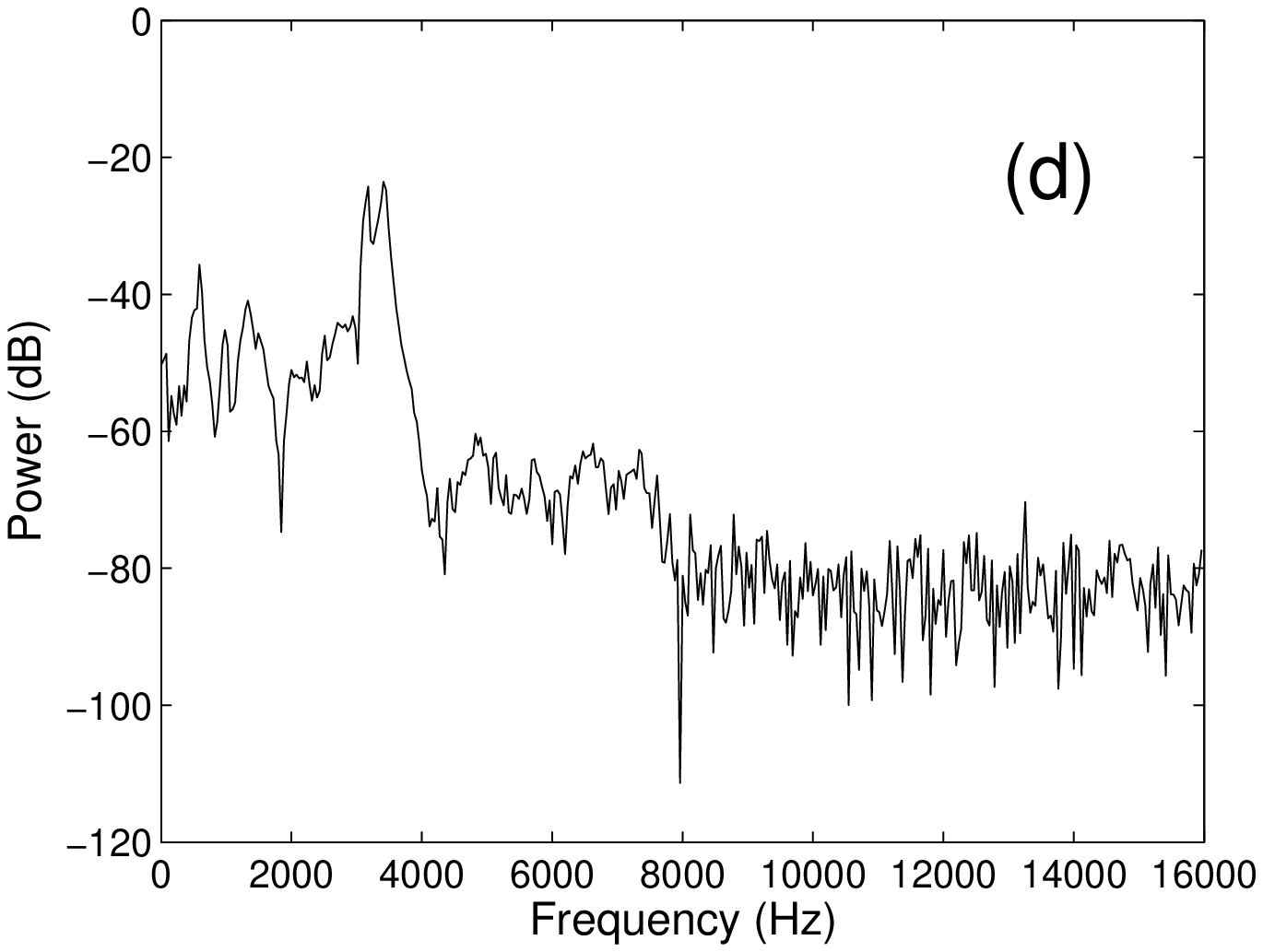}\\
    \includegraphics[width=6 cm]{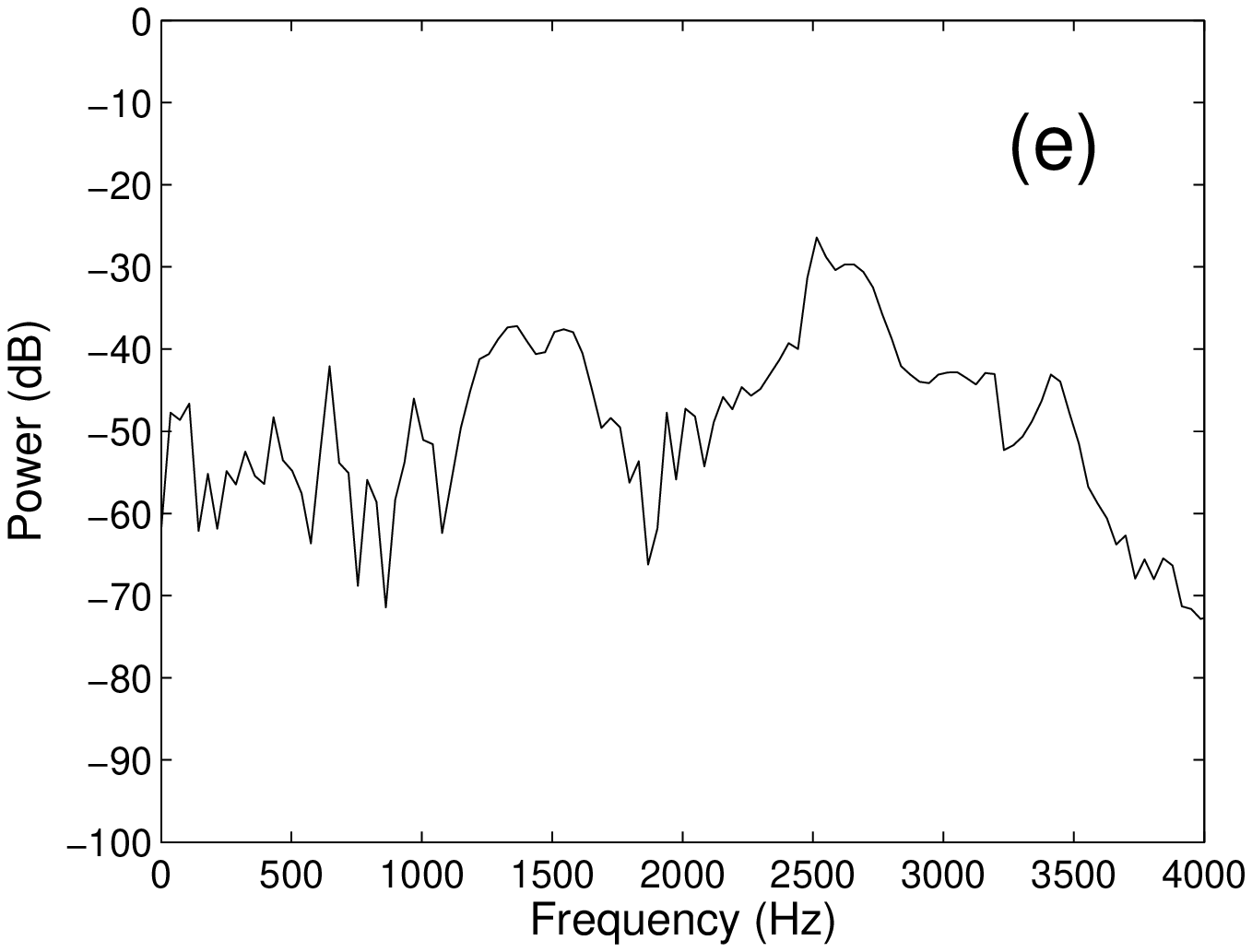}
    \includegraphics[width=6 cm]{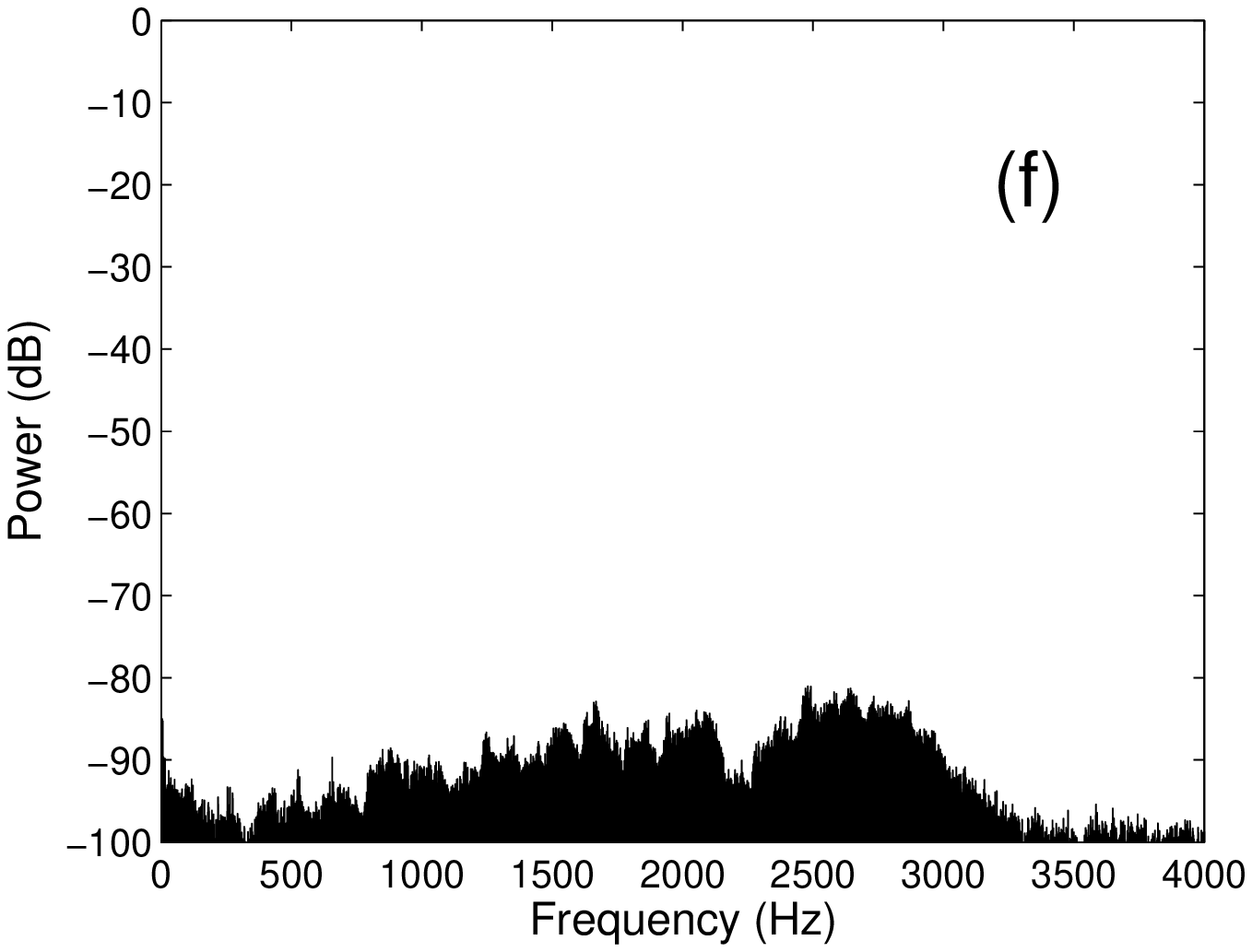}\\
    \caption{a partial gallery of the PSDs of a "true nucleation" event (a), in comparison with typical gel-associated background acoustic events (b) fracture, (c) N$_{2}$ release and (d) microleak, as well as typical acoustic backgrounds such as (e) water input bubbles and (f) ventilation system operation. Although the $\tau_{0}$, $\mathcal{F}$ parameters vary with each, the characteristic PSD is preserved.}
\label{fig5}
\end{figure*}

\subsection{Longevity}

The new Phase II instrumentation, together with improved SDD fabrication chemistry, permitted a significant improvement in the usable lifetime of a detector. Phase I SDDs were usable over $\sim$ 40 d of continuous exposure", conservatively adopted for devices fabricated without PVP in which signal avalanches began to appear in the detectors as a result of fractures and their propagation -- which the piezoelectric instrumentation at the time was unable to discriminate from true bubble nucleations \cite{jptese,njp}. The improved instrumentation, with its ability to identify the fracture dynamics, permits the state of the SDD to be monitored during its operation, extending the previous lifetime.

Moreover, the fracturing is slowed by the addition of PVP, which increases the required gel fracturing energy and viscosity, and strengthens the gel matrix while further reducing the already low solubility of the C$_{2}$ClF$_{5}$. The standard fabrication recipe was developed from Phase I studies of the increased SDD lifetime and fracture reduction via use of PVP of various polymerization indices and other additives \cite{tmtese,tomonim}. These studies included control of the droplet size distribution with fractionating times, and neutron and gamma irradiations \cite{tomoap} which explored the rate of fracture occurrence with recipe variations.

In Phase I, all SDD fabrications were made in a Paris laboratory, which necessitated the 700 km overland transport of the SDDs to the LSBB in a state of reduced sensitivity (T $\leq$ 0$^{o}$C, P = 4 bar). Analysis of this transport however indicated fractures and bubble formation (probably of mechanical origin) as well as the formation of clathrate hydrates \cite{tomoap} because of the cooling, which create surfaces for new bubble formation when being warmed to the device operating temperature. In consequence, the fabrication laboratory was transplanted to the LSBB and all Phase II SDD science fabrications made in a 210 mwe underground "white room" $\sim$ 1 km from GESA; "cold storage" was also discontinued to eliminate the formation of clathrate hydrates. These protocol changes provided significant improvements in the detector performances. With "standard" SDDs submerged to the center of an otherwise unshielded 700 liter waterpool and the state of the SDD monitored with the current acoustic instrumentation, a loss of detection stability via fractures occurred only after $\sim$ 100 days of operation; $\leq$ 6 fracture-induced events were recorded per detector prior to stability loss. Fractures naturally resulting from Oswald ripening of the bubbles ultimately led to performance degradation suggested in Ref. \cite{cmt1}, the latter however easily observed with the new instrumentation.

In consequence of the above, SIMPLE SDDs are run to exhaustion with the state of the detector monitored acoustically -- unlike PICASSO which recompresses each 4 days of operation to reliquify the C$_{4}$F$_{10}$ gas bubbles, Furthermore, both the SDD lifetime and acoustic detection efficiency are naturally increased if the device is only weakly irradiated initially \cite{jptese,tmtese}, i.e. the number of bubbles which can grow into fractures is small: given the ability of the instrumentation to identify both fractures and their propagation, science SDDs are not irradiated before or during the measurements, but rather the state of each acoustically-monitored throughout a measurement, and the devices calibrated via weak irradiations afterwards.

\subsection{Nuclear Recoil Events}

\subsubsection{Response}

Phase I neutron irradiations \cite{jptese,itn} indicated an initial overall bubble detection efficiency of 100\%, with a decreasing detector response after $\sim$ 100 bubble formations resulting from sound attenuation because of the increasing bubble population; the decrease is correctable since both the droplet loss and loss per droplet size are exponential (see Ref. \cite{jptese}). Phase II studies similarly yielded an initial detection efficiency of 100\% independent of the bubble location \cite{felizspatloc}. Investigations of the impact of normal small fabrication variations in the SDD fabrications showed no significant changes in the signal parameters identified with the various signal origins \cite{felizfabvar}.

The SRIM-calculated LET for fluorine recoils in C$_{2}$ClF$_{5}$ is shown in Fig. 5: a recoil generally travels $<$ 1 $\mu$m in the liquid with a LET $>$ LET$_{c}$ only over a fraction of the distance, resulting in the formation of O(1) proto-bubbles.

\begin{figure}[h]
  \includegraphics[width=8 cm]{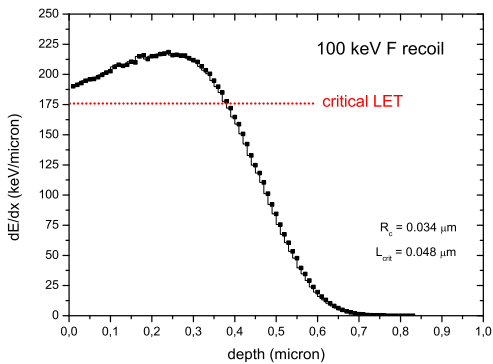} \\
  \caption{ SRIM-calculated LET for 100 keV fluorine recoil ions in C$_{2}$ClF$_{5}$ as a function of penetration depth. The conditions for bubble nucleation are satisfied only over $\sim$ 0.4 $\mu$m of liquid penetration. }
\label{fig3}
\end{figure}

Experimentally, the recoil-generated events are distributed normally when scaled in the acoustic power of the event, ln($\mathcal{A}_{nr}^{2}$), as seen in Fig. 6 with a neutron irradiation-only 900 ml "standard fabrication" device. From Eq. (4), the distribution should reflect the droplet size distribution, as also shown in Fig. 6  with $\mathcal{A} \rightarrow$ gr$^{3}$ and "g" a conversion factor.

\begin{figure}[h]
  \includegraphics[width=8 cm]{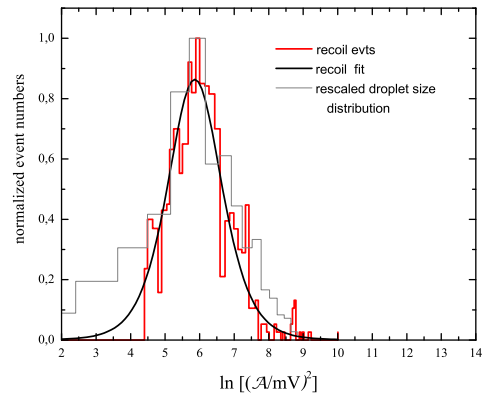}\\
  \caption{amplitude analysis of recoil events generated by neutron irradiation (solid), in comparison with a droplet size distribution (dashed) generated by the "standard fabrication protocol" of Sec. II.A when scaled as $\mathcal{A} \rightarrow$ gr$^{3}$. The Gaussian fit to the irradiation data yields a mean of 5.8, 4.4$\sigma$ below ln($\mathcal{A}^2)$ = 9.2.}
  \label{fig3}
\end{figure}

On the basis of over 45 neutron calibration measurements, a recoil acceptance window was defined with an upper limit of ln($\mathcal{A}_{nr}^{2}$) = 9.2 ($\mathcal{A}_{nr}$ = 100 mV), the highest observed amplitude in any of the measurements and 4.4$\sigma$ above the mean of the Fig. 6 recoil distribution, giving a recoil acceptance of $>$ 99.99\% if a large number of nucleations occurs, slightly better than the conservative "$>$ 97\% acceptance" initially stated in Ref. \cite{prl1}. The various SDDs, with unavoidable small variations in the fabrication protocol \cite{felizvary}, however yielded a range of recoil distribution means 3.5-4.6$\sigma$ below 100 mV: rather than average, the "worst case" 3.5$\sigma$ distance was adopted as a lower limit. Although this gives a recoil containment probability of $>$ 99.98\% with $<$ 0.02\% probability for recording a recoil event above the limit, we however maintain the previous, conservative "$>$ 97\% acceptance" to encompass the "known unknowns" of this measurement aspect.

\begin{figure}[h]
  \includegraphics[width=8 cm]{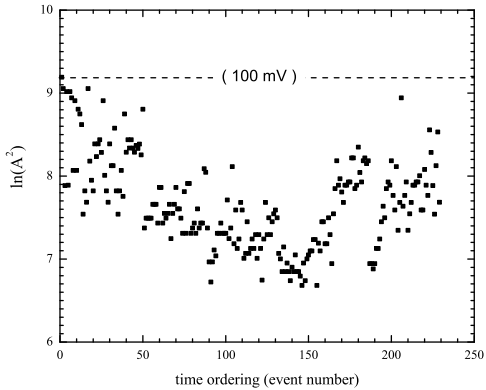}\\
  \caption{Scatter plot of the response-corrected first 250 recorded event amplitudes in the SDD neutron irradiation of Fig. 6, yielding a recoil distribution with mean 3.5$\sigma$ below ln($\mathcal{A})^{2}$ = 9.2, indicating the exposure-dependent populating of the eventual recoil distribution with a bias to largest events first.}
  \label{fig3}
\end{figure}

Although neutrons interact indiscriminately within the droplets, the overall SDD interaction should be preferentially biased to largest droplets first as a result of geometric cross sections, and the acceptance window should be correspondingly populated in time. Fig. 7 displays the response-corrected first 250 recorded recoil event amplitudes of the "worst case" 3.5$\sigma$ irradiation, indicating the populating of what eventually becomes a Gaussian distribution; as anticipated, the earliest events are predominantly of large $\mathcal{A}$, with no signal amplitudes above ln($\mathcal{A}^{2}$) = 9.2 of the first event. Given this and the number of calibration measurements, with the first 10 events having the approximately the same amplitude in each case, an estimate of the probability that a recoil-generated nucleation during a WIMP search has a signal amplitude over 100 mV is $\sim$ 1/450 = 0.22\%.

\subsubsection{Nucleation Efficiency}

The Seitz theory suggests the efficiency of a bubble nucleation for E$_{R}$ = E$_{thr}$ to equal 1. The bubble nucleation efficiency $\varepsilon^{A}$ of an ion of mass number A recoiling with energy E$_{R}$ however depends on the statistical nature of the energy deposition and its conversion into heat \cite{barnabe}:

\begin{equation}\label{vmin}
    \epsilon^{A} =  1 - exp[-\Gamma(1-E_{thr}^{A}/E_{R})] ,
\end{equation}

\noindent with $\Gamma$ a detector-dependent parameter characterizing the slope of the response curve above E$_{thr}^{A}$.

Monochromatic neutron irradiation data \cite{itn}, obtained as a function of temperature at fixed pressure and previously analyzed via Seitz theory, were re-analyzed using Eq. (5) with the reaction rates,

\begin{equation}\label{reac}
    R(E_{n}) =  \phi(E_{n})V \sum_{A,j} \sigma^{A}_{j}(E_{n}) N^{A} \epsilon^{A},
\end{equation}

\noindent where V is the liquid volume, $\sigma^{A}_{j}$ is the A-specific reaction cross section of type j (from ENDF60), $\phi$(E$_{n}$) is the flux of neutrons with energy $E_n$ incident on the droplet, and N$^{A}$ is the atomic density of the A$^{th}$ species of the liquid. The $\phi$(E$_{n}$) included scattering in both the gel and the water bath. The E$_{thr}^{A}$ corresponding to temperature was computed via Eqs. (1)-(3) using $\Lambda$ = 1.40, and then identified with the maximum recoil energy E$^{max}_{R}$ = [4A/(1+A)$^{2}$]E$_{n}$.

\begin{figure}[h]
  \includegraphics[width=8 cm]{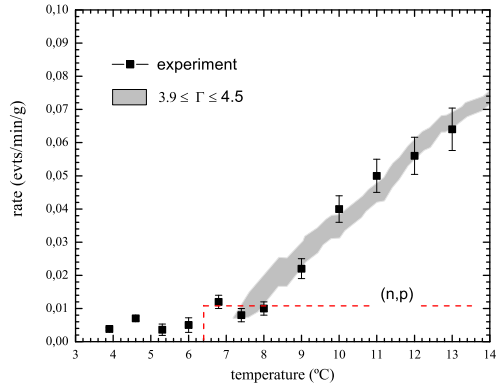}\\
  \caption{a typical re-analysis of monochromatic neutron (54 and 149 keV) SDD irradiation data at 2 bar, including all neutron-liquid interactions weighted by Eq. (7) with $\Gamma$ a free parameter; the shaded region corresponds to H(E$_{c}$) with $\Gamma$ = 4.2 $\pm$ 0.3.}
  \label{fig3}
\end{figure}

Since a SDD is a threshold device, the experimental temperature spectra correspond to the integral spectra of the reversed differential response, or

\begin{equation}\label{reac}
    H(E_{c}) =  \int_{0}^{\infty} R(E_{n})dE_{n}.
\end{equation}

\noindent A typical reanalysis is shown in Fig. 8 for 54 and 149 keV neutrons on a C$_{2}$ClF$_{5}$ SDD at 2 bar, and includes the (n,p) component resulting from an epithermal leak-through of reactor neutrons in the beam filters: with $\Gamma$ a free parameter, and including all reaction contributions for each target nuclide, a best fit yielded $\Gamma$ = 4.2 $\pm$ 0.3 as shown by the shaded region of Fig. 8. The same result was obtained for the same irradiations at 1 bar, and with beams of 24 keV at both pressures.

\subsection{Alpha-Recoil Event Discrimination}

Studies of the $\alpha$ response were initiated using separate "standard fabrication protocol" devices $\alpha$-doped with U$_{3}$O$_{8}$, based on a desire for pristine device response and their response separation. The results \cite{prl1}, when compared with those of the neutron irradiations as a function of ln($\mathcal{A}^{2}$), identified a gap of 30 mV between the two distributions, with the $\alpha$ distribution appearing as a truncation of an equally Gaussian-like distribution.

Since the first SDDs used for recoil calibrations differed in volume and cross section (900 ml, rectangular) from those used for $\alpha$ calibrations (250 ml, circular), the same calibration measurements were performed with single 250 ml SDDs, first with neutron irradiations and then with $\alpha$ doping as before. A typical example, shown in Fig. 9 and contained within Ref. \cite{reply2}, reproduces well the separated device recoil results of Ref. \cite{prl1}, with the distribution continuing to reflect the droplet size distribution, and a well-resolved gap separation of 20 mV between the two distributions. The apparent increase in the recoil distribution mean from that of the separate SDD's of Ref. \cite{prl1} is attributed to the smaller size of the SDD in the measurements (which gives a 20\% increased $\mathcal{A}^2$, as observed).

\begin{figure}[h]
  \includegraphics[width=8 cm]{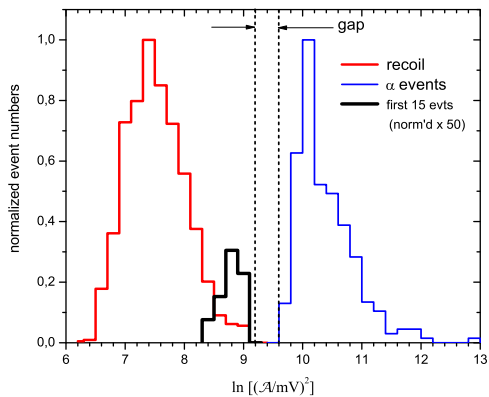}\\
  \caption{experimental results obtained with a single C$_{2}$ClF$_{5}$, standard fabrication detector at 2 bar and 9$^{o}$C, first irradiated with Am/Be neutrons, and then doped with U$_{3}$O$_{8}$ $\alpha$'s, yielding 867 $\alpha$ and 1280 recoil events. The indicated gap corresponds to a separation of 20 mV, which is readily resolved with the 0.3 mV of the acquisition electronics. The asymmetry in the $\alpha$ response is due to the relationship between droplet size distribution and the Bragg curve above LET$_{c}$ of the $\alpha$'s in the liquid, as discussed in the text. The central peak represents the first 15 events of the neutron irradiation, amplified by a factor 50 for visibility.}
  \label{fig3}
\end{figure}

Because calibrations show the $\alpha$ distributions to begin 20-43 mV above the adopted 100 mV cutoff of the neutron distribution, a higher upper limit on the recoil acceptance window could have been defined, with an even smaller expected inefficiency for WIMP-generated events.

Empirically, the existence of the observed gap requires that

\begin{equation}\label{reac}
    \mathcal{A}_{\alpha}^{2}|_{min} >  \mathcal{A}_{nr}^{2}|_{max}.
\end{equation}

\noindent Both $\alpha$ and recoil event distributions derive from the same droplet distribution in the SDD, and overlap of the two could be expected (as observed in Ref. \cite{newinsite}). On the other hand, the $\alpha$ dE/dx \textit{a priori} differs significantly from that of the recoil ion, as shown in the SRIM-calculated energy loss of 5.5 MeV $\alpha$'s in C$_{2}$ClF$_{5}$ of Fig. 10, where for simplicity an $\alpha$ origin at the droplet surface is assumed (since the $\alpha$-emitters tend to migrate to the droplet surfaces because of the actinide complex ion polarity). Within this model, an $\alpha$ achieves LET $>$ LET$_{c}$ only between 32-40 $\mu$m of liquid penetration: r $\leq$ 16 $\mu$m constitutes a lower size cutoff (r$_{co}$) to the droplet participation - the droplets cannot contribute to bubble nucleation since the $\alpha$ transits the droplet without achieving LET$_{c}$. Droplets with 16.5 $<$ r $<$ 19.5 $\mu$m (including a 0.5 $\mu$m following Fig. 5 to provide a formation distance since the fluorine recoil generates O(1) proto-bubbles), support multiple proto-bubble formation for which a SRIM-estimate, neglecting statistical effects near the two cutoffs, gives the number (n$_{pb}$) of proto-bubbles formed over the $\alpha$ path length in the liquid above LET$_{c}$ as  $\sim$ 169 for the full 8 $\mu$m path length, or n$_{pb} \sim$ 22/$\mu$m.

\begin{figure}
\includegraphics[width=9 cm]{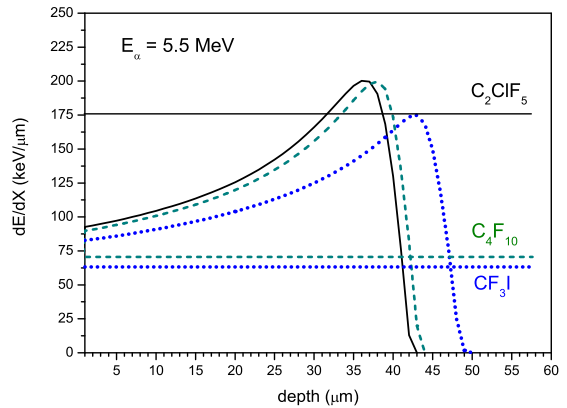}\\
  \caption{Bragg curves for 5.5 MeV $\alpha$'s in C$_{2}$ClF$_{5}$ (solid), C$_{4}$F$_{10}$ (dashed) and CF$_{3}$I (blue dotted) as a function of penetration depth, with the LET$_{c}$ of each at the operating temperatures and pressures indicated in Refs. \cite{prl2,pic2012,coupp2012}. The Bragg peak in C$_{2}$ClF$_{5}$ above the LET$_{c}$ of 176 keV/$\mu$m is achieved with 32-40 $\mu$m penetration of the liquid. The $\Lambda$ for the C$_{4}$F$_{10}$ and CF$_{3}$I are obtained from the phenomenological $\Lambda$ = 4.3($\rho_{v}$/$\rho_{l}$)$^{1/3}$, which is not verified for either; both \cite{coupp2012,pic2012} however claim larger $\Lambda$, which reduces the respective LET$_{c}$ relative to that indicated.}
\label{fig3}
\end{figure}

This is however weighted by the "containment" probability that the required $\alpha$ trajectory above LET$_{c}$ lies within a droplet, since for a 33 $\mu$m droplet diameter, the $\alpha$ trajectory must be a diameter; as the droplet diameter increases, the spherical cone defined by r$_{co}$ opens and the containment phase space increases. This is shown in Fig. 11, where the ratio of the cone-to-droplet volume is shown as a function of droplet diameter: the containment probability rises from near zero at 2r = 32 $\mu$m to $\sim$ 60\% at 2r = 40 $\mu$m; only droplets with 2r $\geq$ 60 $\mu$m have a 90\% containment probability. For trajectories longer than 40 $\mu$m within a droplet, no further proto-bubble formation occurs since the LET is below LET$_{c}$.

For higher E$_{\alpha}$, the Bragg peaks of Fig. 10 are translated to larger penetration depths, with the gap criterion more easily satisfied since r$_{co}$ increases, as also the Bragg width above LET$_{c}$.

Integration of Eq. (4) over the droplet evaporation time gives the energy W released in the event, W = $\mathcal{A}^2\tau$, which is the same for same-sized droplets independent of the nucleation stimulus. Together with the above size considerations, Eq. (8) becomes

\begin{equation}\label{reac}
    r_{co}^{6} \tau_{\alpha}^{-1} > r_{max}^{6} \tau_{nr}^{-1}.
\end{equation}

\noindent Since each proto-bubble constitutes a center for the droplet evaporation, and formation occurs on ns time scales, well below the instrumental resolving time, droplet evaporation with multiple proto-bubble formation is correspondingly accelerated as $\tau_{\alpha}$ $\rightarrow$ $\tau_{nr}$/n$_{pb}$ with $\tau_{nr}$ the droplet evaporation time for a single proto-bubble, and Eq. (9) reduces to r$_{co}^{6}$n$_{pb} >$ r$_{max}^{6}$.

\begin{figure}
\includegraphics[width=8 cm]{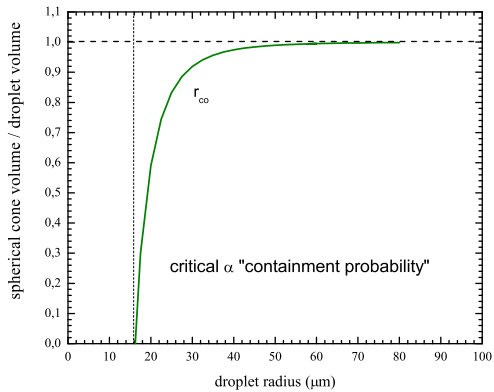}\\
  \caption{Phase space estimate of the containment probability that a droplet of diameter 2r contains an $\alpha$ trajectory above LET$_{c}$, as described in the text.}
\label{fig3}
\end{figure}

The power ratio of the $\alpha$-to-recoil events is

\begin{equation}\label{reac}
    \frac{\mathcal{A}_{co}^{2}}{\mathcal{A}_{nr}^{2}} \sim  \frac{r_{co}^{6}}{r_{nr}^{6}}n_{pb},
\end{equation}

\noindent yielding a ratio of $\sim$ 22 with the above SRIM estimate, as observed in Fig. 9 and below the 400 of Ref. \cite{prl1} as a result of the smaller SDD volume.

The r$_{co}$, and the trajectory containment probability are responsible for the asymmetry of the ln($\mathcal{A}_{\alpha}^{2}$) distribution in Fig. 9. Adjustment of the fractionating time/speed in the SDD fabrication process so that the droplet size distribution $<$r$>$ $\sim$ 2r$_{co}$ places the minimum of the $\mathcal{A}_{\alpha}^{2}$ distribution distinguishably above the maximum of the $\mathcal{A}_{nr}^{2}$ distribution. The "standard fabrication protocol" described in Sec. II.A was in consequence adopted to result in reproducible, homogeneous and approximately normally-distributed droplets with $<r>$ = 30$\pm$7.5 $\mu$m, near the Bragg peak of the $\alpha$ dE/dx, as seen in Fig. 12 from optical microscopy of batch samples in Phase I-II studies; the fraction of droplets with r $\geq$ 80 $\mu$m = 0.

\begin{figure}[h]
  \includegraphics[width=8 cm]{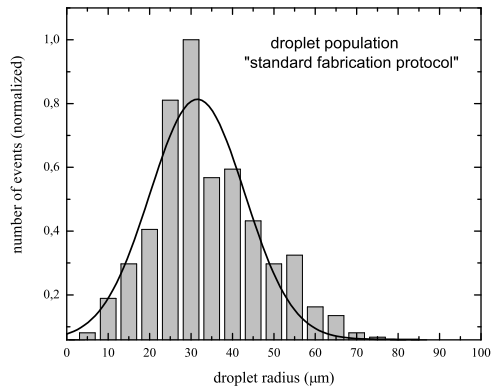}\\
  \caption{sample distribution of 522 C$_{2}$ClF$_{5}$ droplet sizes taken from the "standard fabrication protocol" of Sec. II.A.2, together with a Gaussian fit to the distribution with $\chi_{r}^{2}$ = 4.6.}
\label{fig2}
\end{figure}

Fig. 10 also indicates the $\alpha$ dE/dx curves for the liquids of PICASSO and COUPP (C$_{4}$F$_{10}$ and CF$_{3}$I), with LET$_{c}$ of 70 and 63 keV/$\mu$m at their respective operating temperatures and 1 bar pressures \cite{pic2012,coupp2012}. These are \textbf{\textit{calculated}} with the phenomenological $\Lambda$ = 4.3($\rho_{v}$/$\rho_{l}$)$^{1/3}$, which is not verified for either as stated above; both in fact claim larger $\Lambda$, which reduces accordingly the respective LET$_{c}$ relative to that shown. As seen, the $\alpha$ dE/dx in both cases exceed their respective LET$_{c}$'s over the majority of the $\alpha$-trajectory, eliminating any r$_{co}$ and implying via Eq. (8) a serious overlap of the recoil and $\alpha$ event amplitudes. Single SIMPLE SDD calibrations \cite{inprog}, similar to those of the Fig. 9 studies but with separately a 100$\times$ stiffer gel and a 2$\times$ larger droplet size distribution, indicate for both a further shift of the recoil distribution to higher amplitudes, with the $\alpha$ "peak" fixed but an increasing low ln($\mathcal{A}_{\alpha}^{2}$) tail, a merging of the two distributions with loss of the gap, and a reduction in power ratio.

\section{PHASE II EXPERIMENTAL SETUP}

The search experiment was conducted in the GESA cavern of the LSBB, located 505 m beneath the "Grand Montaigne" of the Plateau d'Albion near Apt in southern France. The rock is karstic, with a primary carbonate composition and numerous water reservoirs distributed within the mountain and surrounding the cavern \cite{lsbb}.

GESA is shielded from the surrounding rock by 30-100 cm of low grade concrete dating from 1971, which is internally sheathed by a 1 cm thickness of steel to provide a Faraday cage with EM fluctuations of order 2 fTesla$\sqrt{Hz}$ above 40 Hz.

\begin{table*}
\caption{\label{table} Active liquid mass in each of the SDDs in the two stages of the Phase II measurement.}
\begin{ruledtabular}
\begin{tabular}{ccccccccccccccccc}
  position & A & B & C & D & E & F & G & H & I & J & K & L & M & N & O & P    \\
  \hline
   Stage 1 mass (g) & 13.2 & 12.0 & 11.2 & 12.1 & 21.2 & 10.6 & 18.8 & 17.7 & 16.5 & 17.8 & 12.3 & 12.4 & 10.2 & 14.4 & 8.4 & 0.0  \\
  \hline
  Stage 2 mass (g) & 12.8 & 18.6 & 15.9 & 14.1 & 13.7 & 12.2 & 12.9 & 11.0 & 15.4 & 14.8 & 11.5 & 14.3 & 15.2 & 17.2 & 15.6 & 0.0  \\
\end{tabular}
\end{ruledtabular}
\end{table*}

\subsection{Installation}

A schematic (not to scale) of the experimental setup within GESA is shown in Fig. 13. The measurements were conducted in two Stages, each of 15 SDDs distributed as a planar array in alternating positions of a 16 cm square lattice, with a 10 cm separation from one another. Each SDD contained between 8-21 g of C$_{2}$ClF$_{5}$ as shown in Table II, comprising a total active mass of 0.208 kg (Stage 1) and 0.215 kg (Stage 2). Each array was submerged to 50 cm above the bottom of a 700 liter water pool.

The Stage I water pool rested on a dual vibration absorber placed atop a 20 cm thick wood pedestal resting on a 50 cm thick concrete floor; the pool was surrounded by three layers of sound and thermal insulation. Between the two Stages, the water pool was raised to accommodate an additional 10 cm of high density polyethylene; an additional 10 cm of wood and paraffin bricks were added to gaps in the wood pedestal support to bring the thickness to 30 cm.

\begin{figure}[h]
  \includegraphics[width=6 cm]{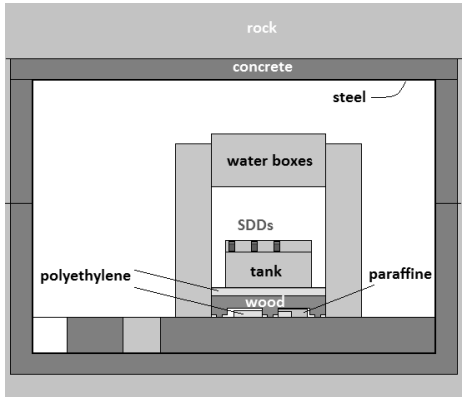}\\
  \caption{schematic (not to scale) of the Phase II experiment Stage 2 disposition in the GESA facility
  of the LSBB. }
  \label{fig1}
\end{figure}

The temperature of the waterpool was set to 9.0 $^{o}$C, monitored with a stainless steel pt100 probe (H62IKA) and controlled to 0.1 $^{o}$C by a polycryothermostat (Huber UC025-H) located 5 m outside GESA. Pool connections were made with thermally-insulated plastic tubing fed through the GESA wall.

A 50-75 cm thick outer water shielding, constructed of 20 liter water boxes and locally-available tap water, surrounded the insulated pool and pedestal both laterally and above. Between Stage 1 and 2, the outer water shield was rebuilt to further eliminate neutron thruways.

Additionally, the Phase I acoustic baffling of the GESA venting was removed and the air circulated at 0.2 m/s, purging the cavern air $\sim$ 10 times per day to reduce the ambient radon levels to $\leq$ 100 Bq/m$^{3}$ during the two Stages. The water pool covered the detectors to 6 cm above the active detector material, and was circulated at 25 liter/min, with the entry water at the pool surface to replace the top 1 cm layer of the water pool each minute in an effort to reduce the water radon concentration via atmospheric diffusion; the increased accompanying acoustic background was easily identified and rejected following a Sec. III.B analysis (see Figs. 4(e) and 4(f)). Nearby work interventions in the LSBB during Stage 2 however forced cessations in the air circulation for few-day periods, increasing the local radon levels by as much as 30$\times$ - which were however logged on the LSBB intranet throughout for later analysis of the variations in SDD $\alpha$ response during the measurements.

During SDD installations, each was pressurized to 2.00 $\pm$ 0.05 bar, with its pressure channel recorded in a 1 Terabyte pc at 1 kSps, located 7m outside GESA. Detector microphone outputs were recorded in one of two similarly located TB PC's, each via a shielded telecommunications-grade cable (SHC68-68-EPM) link which eliminated pickup resulting from cable motion, and reduced pickup from ambient acoustic noise and site activity by a factor 10 even when exaggerated: the noise level was $\sim$ 2-3 mV per acoustic channel \cite{prl1}. The minimum voltage accuracy was 52 $\mu$V with a sensitivity of 6 $\mu$V. The DAQ acquisition rate was set at 8 kSps in a differential mode, and recorded in time-tagged files of 8 min each in a Matlab platform; time resolution was $\sim$ 125 $\mu$s.

In Stage 1, the DAQ was initiated with each SDD installation during the 18 day setup period, which recorded the interventions and provided "state-of-the-detector" calibrations as well as additions to the acoustic background "gallery" of the detector/DAQ operation over the installation period. In Stage 2, the DAQ was initiated only after the installation of each 8 detector array was completed. In both cases, with the shield partially removed, particle-induced events were recorded. Also in both Stages, an additional, similarly-installed, freon-less but otherwise identical SDD, served as an acoustic veto.

\subsection{Backgrounds}

The cavern is characterized by a muon thru-flux of 2.0 $\times$ 10$^{-2}$ cm$^{-2}$s$^{-1}$, with the ambient neutron flux due to the surrounding rock at 1500 mwe well-below 4 $\times$ 10$^{-5}$ n/cm$^{2}$s \cite{waysand}. Radio-assays of the U/Th content of the shielding concrete yielded 1.90 $\pm$ 0.05 ppm $^{232}$Th and 0.850 $\pm$ 0.081 ppm $^{238}$U; of the steel, 3.20 $\pm$ 0.25 ppb $^{232}$Th and 2.9 $\pm$ 0.2 ppb $^{238}$U. These obtained from chemical analyses of the rock and concrete composition, and $\gamma$ spectroscopy. The concrete levels are generally at the same levels as those recorded in deeper underground locations such as Canfranc, Modane, and Gran Sasso \cite{canfranc,modane,gransasso}. Measured radon levels, without suppression efforts, varied seasonally 28-3000 Bq/m$^{3}$ as a result of water circulation in the mountain during Summer months.

Diffusion of the environmental radon into a detector was suppressed by circulation of the surrounding pool water: with a diffusion length of 2.2 cm, the steady-state radon concentration was reduced by $\sim$ factor 10 at 6 cm below the water surface. The diffusion was also low because of the short ($<$0.7 mm) radon diffusion lengths of the SDD construction materials (glass, plastics). The measured U/Th contamination of the glass was at a level significantly higher than that of the gel (as discovered only following Stage 2, as a result of a supplier-substitution of borosilicate glass in the order). The N$_{2}$ 1 bar over-pressuring of each device inhibits the advective influx of water-born Rn through the device capping, as well as its diffusion from the walls of the glass container into the gel (via stiffening of the gel).

Table III displays the radio-assays of the U/Th content of the dominant shielding and detector construction materials, obtained from chemical analyses of compositions, ion beam analysis of the hydrogen content of the wood and boron in the glass (elastic recoil detection and nuclear reaction analysis + elastic backscattering, respectively), $\gamma$ (rock, concrete, steel, gel) and $\alpha$ (water, wood) spectroscopy, and comparative neutron activation analysis (glass, polyethylene, paraffin). Standard compositions were used in simulations for the remaining materials. The densities of most materials, necessary to the transport codes, were measured. The presence of U/Th contaminations in the gel was measured at $\leq$ 0.1 ppb by low-level $\alpha$ and $\beta$ spectroscopy of the production gel. Unlisted materials contributed negligibly, based on conservative estimates of their U/Th content, and mass and distance from the SDDs.

\begin{table}[h]
\caption{\label{table}materials' radio-assay results and estimated neutron-induced background measurement contributions as described in Ref. \cite {acf}.}
\begin{ruledtabular}
\begin{tabular}{ccccccccccccccccc}
  & $^{232}$Th & $^{238}$U  & Stage 1 & Stage 2  \\
  & (Bq/kg) & (Bq/kg) & (evt/kgd) & (evt/kgd)\\
  \hline
   rock & 0.16 & 5.0 & - & - \\
   concrete & 7.7 & 10.5  & 0.645*  & 2.89$\times$10$^{-3}$\\
   steel & 1.3$\times$10$^{-2}$  & 3.6$\times$10$^{-2}$ & $\sim$ 0 & $\sim$ 0 \\
   water & 5.0$\times$10$^{-5}$ & 3.2$\times$10$^{-2}$ & 2.54$\times$10$^{-3}$ & 2.53$\times$10$^{-3}$\\
   wood & 3.0$\times$10$^{-3}$ & 1.1$\times$10$^{-1}$ & 2.17$\times$10$^{-5}$ & 6.68$\times$10$^{-6}$ \\
   glass & 1.27 & 2.74 & 3.27$\times$10$^{-1}$ & 3.28$\times$10$^{-1}$ \\
   gel & 1.73$\times$10$^{-6}$ & 1.11$\times$10$^{-5}$  & 5.38$\times$10$^{-4}$ & 5.38$\times$10$^{-4}$ \\
   PE/paraffin & $<$ 0.41 & $<$ 0.25 & 0 & 8.44$\times$10$^{-5}$ \\
\end{tabular}
\end{ruledtabular}
\noindent * with shielding gaps
\end{table}

All materials' radio-assays were used as inputs to Monte-Carlo simulations (conducted independently and "in the blind" relative to the signal analyses) of the expected neutron-generated detector event rate. The simulations accounted for the full experimental geometries, and all neutron-C$_{2}$ClF$_{5}$ interactions, including spontaneous fission and decay-induced ($\alpha$,n) reactions. The fission parameters, spontaneous fission probabilities and neutron multiplicities were obtained from Ref. \cite{shores}; spectra and yields of ($\alpha$,n) neutrons originating from each material were obtained from Ref. \cite{mei} assuming secular equilibrium within the decay series. The reaction cross sections were obtained from the ENDF/B-VII.1 library \cite{chadwick}. The initial results were used to define a baseline for the initial shield construct of Fig. 13, to guide the inner shield constructions, and as a guide for the shield reconstruction between the two Stages; these indicated negligible variations for concrete thicknesses of 20-60 cm \cite{acf}.

\begin{figure*}
\includegraphics[width=8.5 cm]{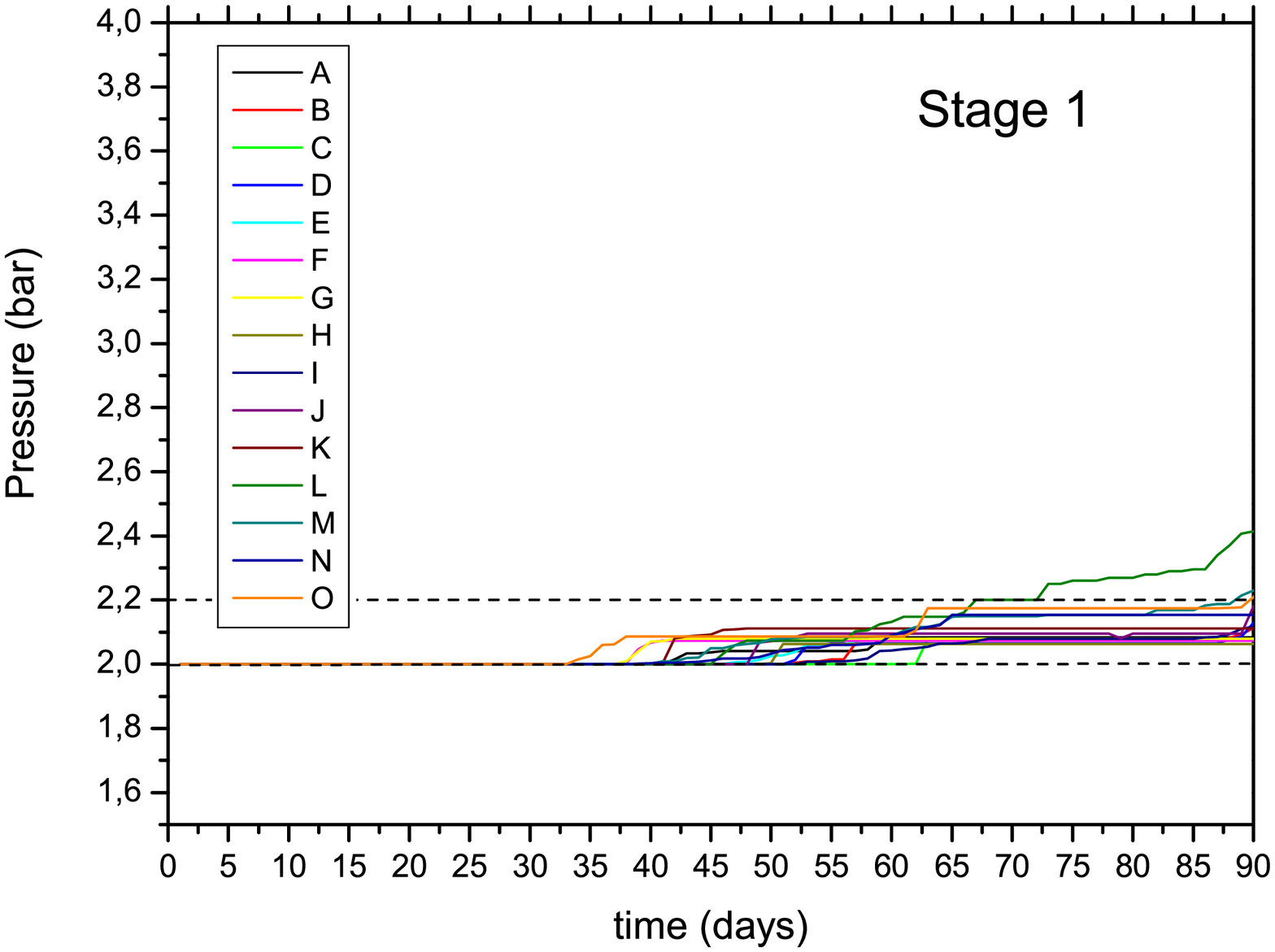}
\includegraphics[width=8.5 cm]{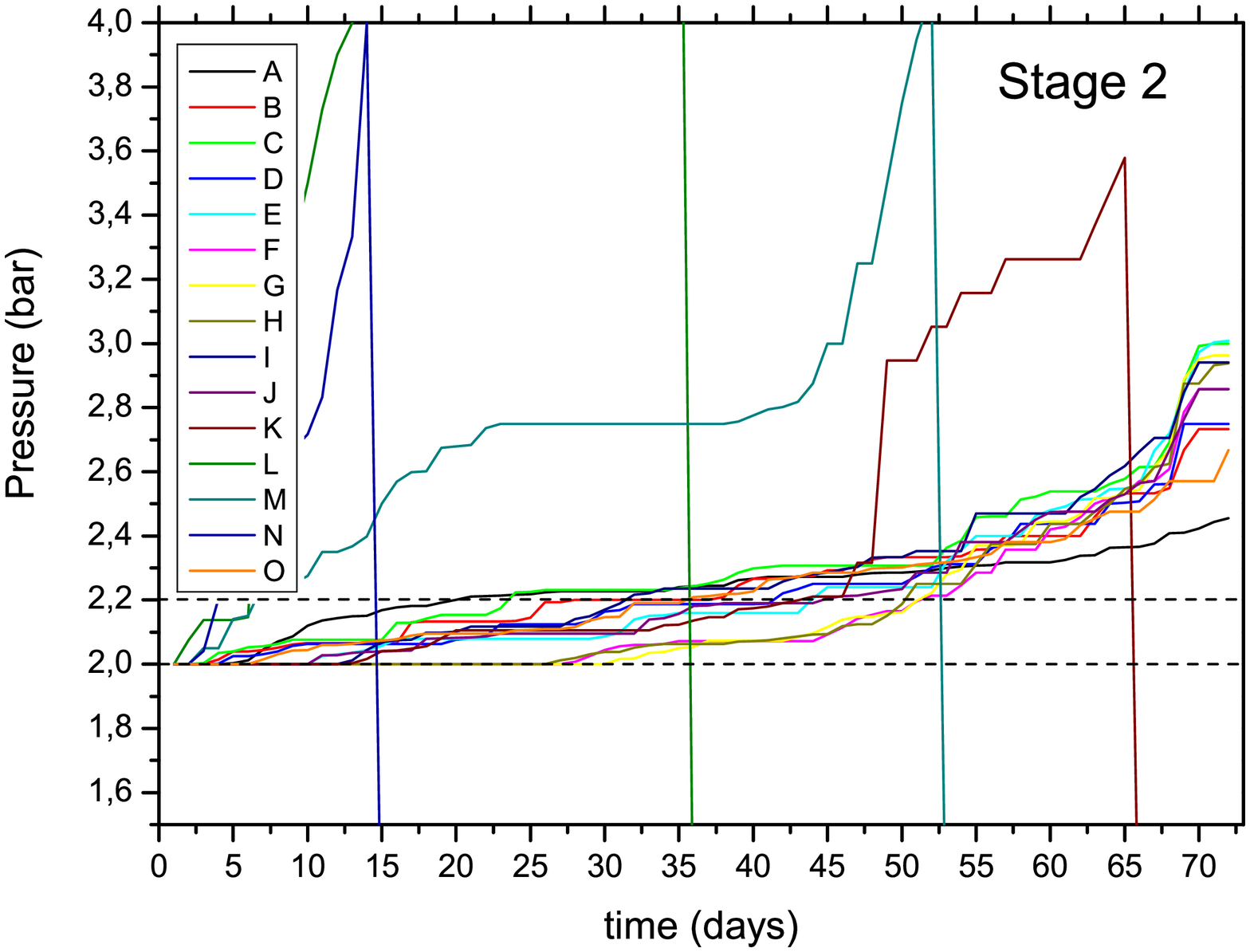}\\
\caption{pressure evolution of the two SDD sets over each of Stage 1 and 2 measurements (same scale), with the dotted lines indicating the 2.00-2.15 bar region. The letter identifications in each correspond to the SDD. The pressures of the SDDs in Stage 2 were slowly increased to above 3 bar, permitting an \textit{in situ} measurement of \textit{$\Lambda$}. The failure of detectors I,L,M in Stage 2 occurred early in the measurement, and the records were not used in the analysis; K ran until day 48.}
\label{fig1}
\end{figure*}

\section{Data Acquisition and Selection}

Following installation of each SDD array, the shield was sealed and science measurements conducted between 14 November 2009 - 05 February 2010 (Stage 1) and 01 May - 22 July 2010 (Stage 2), respectively. Once initialized, the science run signals, temperatures and pressures of each SDD were monitored continuously throughout the measurement period. Radon levels, read by two sensors (Ramon 2.2) located inside and outside the shield, were recorded on a daily basis; in Stage 2, this was augmented by an additional monitor (Durridge RAD7) located outside the shielding and archived continuously on the LSBB intranet.

The waterpool temperature was maintained at 9.0 $\pm$ 0.1 $^{o}$C; a webcam fixed on the Huber temperature readout allowed remote monitoring of the pool temperature on a 24/7 basis via the LSBB intranet. Huber failure due to power loss in both Stages required a manual reset intervention: in Stage 1, a 4.70 kgd measurement loss resulted from weather-induced power failures and consequent intervention delays. In Stage 2, each pc was supported by an uninterrupted power supply: no weather-induced system failures however occurred.

In the last third of the Stage 2 science measurements, the SDD pressures were gradually increased to 3-4 bar with the pool temperature maintained at 9.0$\pm$0.1 $^{o}$C, for the purpose of an \textit{in situ} determination of the nucleation parameter $\Lambda$ of the C$_{2}$ClF$_{5}$.

\subsection{Recoil Threshold Determination}

Operation of a C$_{2}$ClF$_{5}$ SDD at 9$^{o}$C and 2 bar corresponds to a reduced superheat s = 0.34, sufficiently below threshold for all minimum ionizing contributions to be neglected. No indications of $\gamma$'s $>$ 6 MeV were observed.

The pressure recordings of the SDDs in each Stage are shown in Figs. 14. As seen in the Stage 2 record, four SDDs (I,K,L,M) exceeded the containment tolerance and self-destructed; the first three occurred within days of the run initiation and their data is excluded from the science analysis.

The pressure-correlated Stage 2 signal records were first analyzed following the protocol described in Sec. III.B to determine the pressure of $\alpha$ event disappearance, and compared with detailed calculations of E$_{thr}^{\alpha}$ with $\Lambda$ a free parameter. These yielded $\Lambda$ = 1.40 $\pm$ 0.05 with the uncertainty obtained from the temperature and pressure fluctuations, in agreement with the previous determination \cite{jptese}. This calibration further indicated 2.2 $\pm$ 0.05 bar to correspond to E$_{thr}^{nr}$ = 8.0 $\pm$ 0.1 keV; all records with pressures $>$ 2.2 bar were then excluded from further analysis, giving together with the loss of the three SDDs a science exposure of 6.71 kgd.

Stage 1 signal records were similarly pressure-correlated for E$_{thr}^{nr}$ = 8.0 keV and those with pressures $>$ 2.20 bar excluded, reducing the initially-reported 14.1 kgd exposure by \textbf{\textit{0.63}} kgd to 13.47 kgd \cite{prl2}; this is further reduced by 1.94 kgd to 11.53 kgd following the discovery and removal of the previously included 18-day Stage 1 installation data.

\subsection{Signal Selection}

Each recorded data file was subjected to the analysis protocol described in Sec. III.B. Following signal validation and prior to signal characterization, the remaining data set was cross-correlated in time between all SDDs in an array, and coincidences within the system resolving time rejected as local noise events on the basis that a WIMP interacts with no more than one of the in-bath detectors. Figures 15 display the scatter plots of all recorded events of each Stage in terms of $\tau_{0}$ and $\mathcal{F}$, with the boxed areas containing those events identified with the true nucleation event signal parameters following Table I; the absence of acoustic background signal in the "true event" window reflects the differing nature of the event origins as seen in Table I.

\begin{figure*}
    \includegraphics[width=8.5 cm]{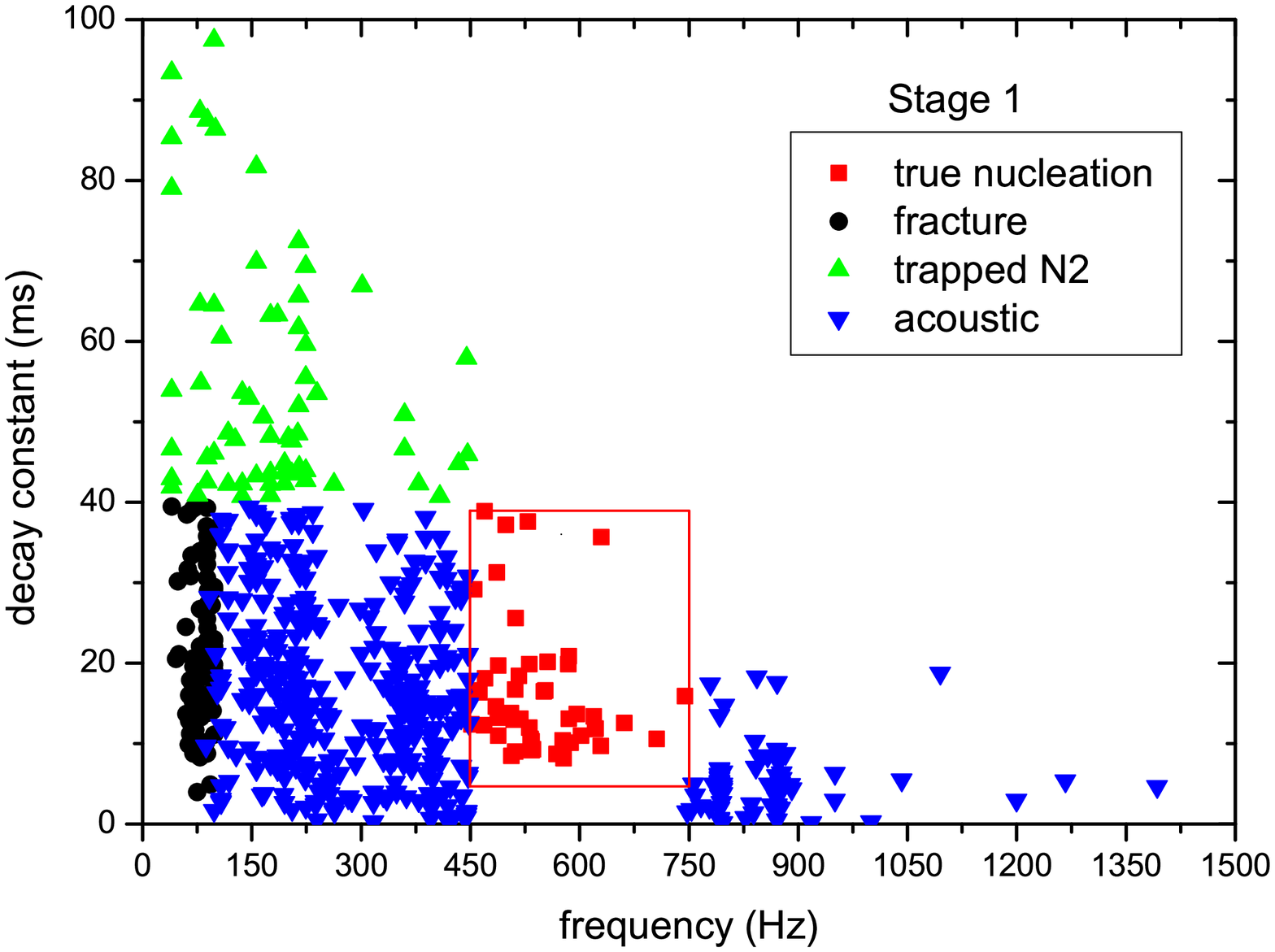}
    \includegraphics[width=8.5 cm]{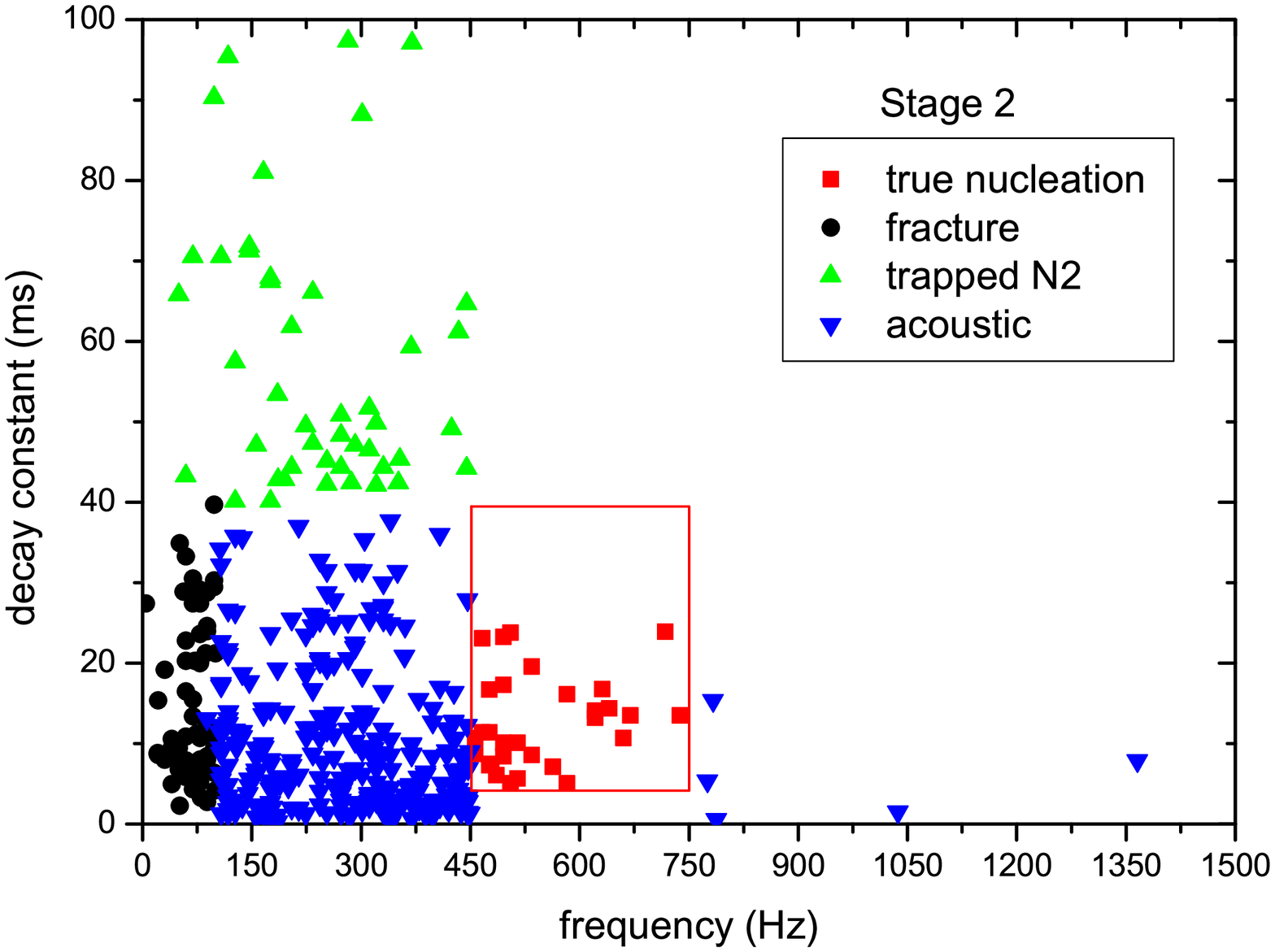}\\
  \caption{scatter plots of the corrected Stage 1 (left) and 2 (right) single event signals with respect to decay constants and frequency. The indicated boxes isolate the events exhibiting particle-induced characteristics. Identification of the event origins follows from comparison of their PSD structures with those of calibration templates. }
  \label{fig3}
\end{figure*}

The PSD structure of each event, as well as those lying near the box borders, was then examined for concordance with the true nucleation event template of Fig. 4(a), which distinguished border events as being either true nucleations or acoustic backgrounds. Figures 16 display similar scatter plots of all PSD-verified signal $\mathcal{A}$ and their respective $\mathcal{F}$, which provide the basis for the recoil event identification. As in Fig. 15, the particle-induced events are separated from the other signals, consistent with the frequency separations of Table I as a result of their differing origins, without leakage of the gel-associated and environmental background signals.

\begin{figure*}
  \includegraphics[width=8.5 cm]{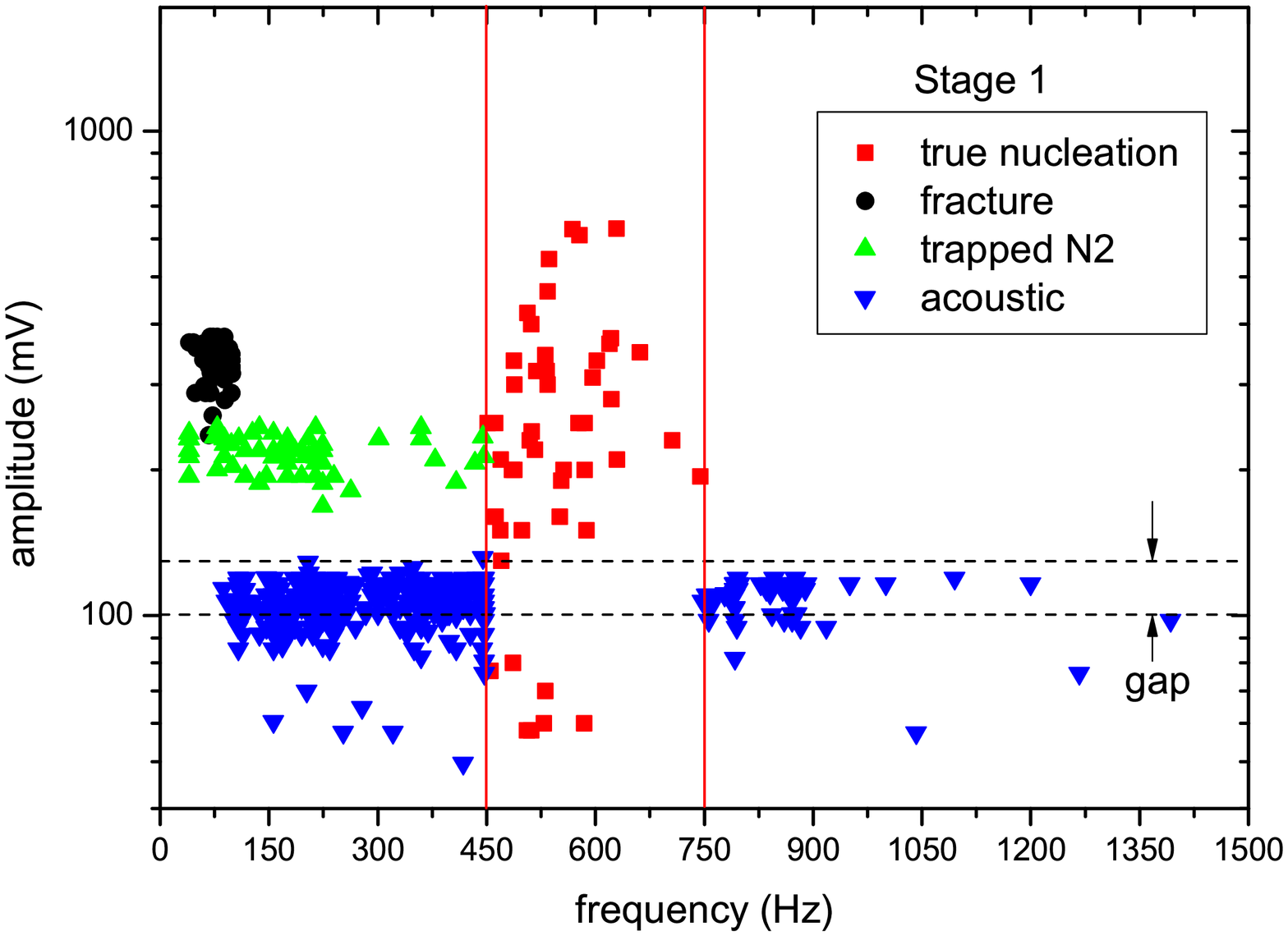}
  \includegraphics[width=8.5 cm]{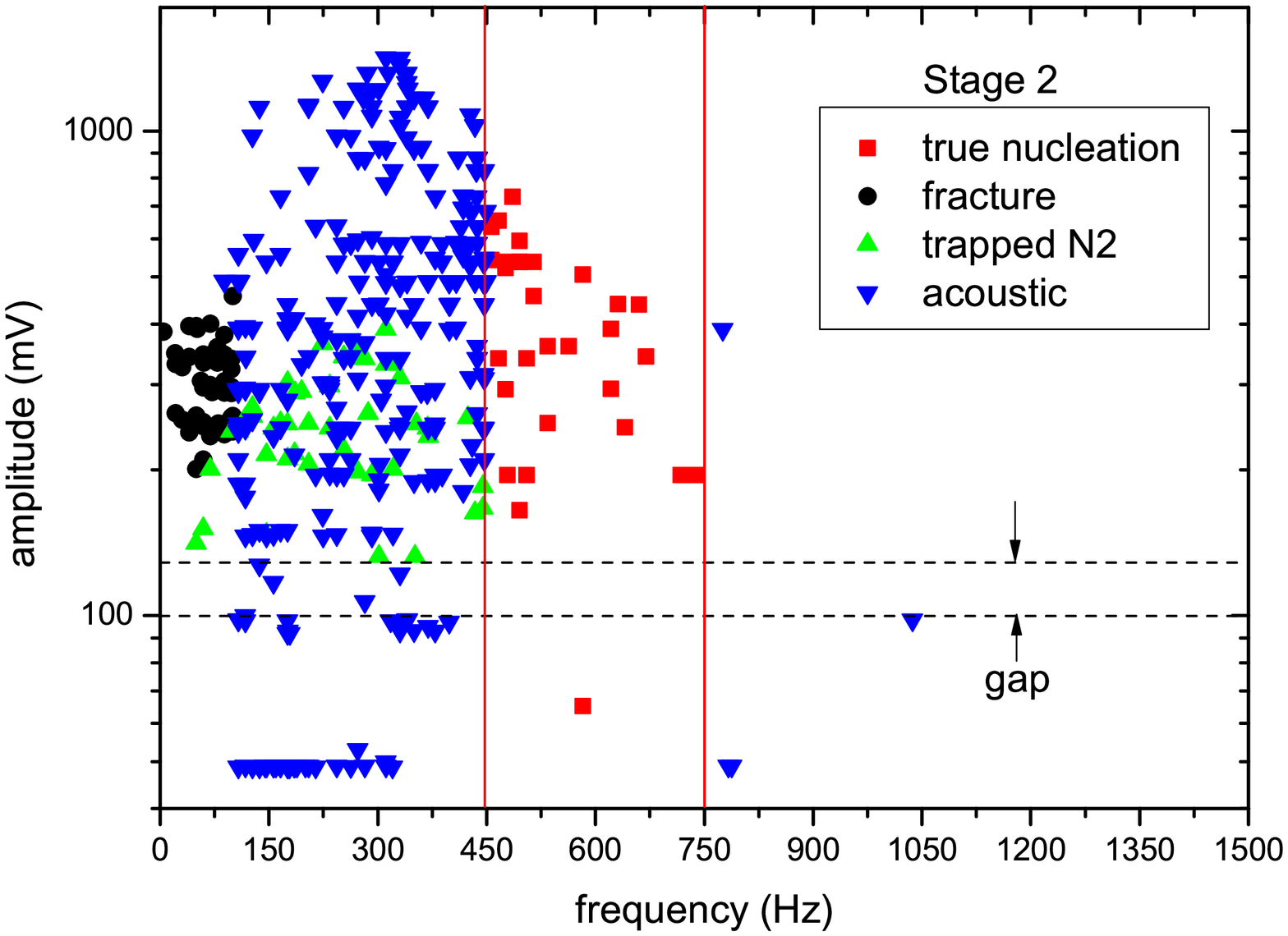}\\
  \caption{scatter plots of the corrected Stage 1 (left) and 2 (right) single event signals with respect to $\mathcal{A}$ and $\mathcal{F}$. Identification of the event origins follows from comparison of their PSD structures with those of calibration templates. }
  \label{fig3}
\end{figure*}

The differences between the event distributions in the two Stages, in particular in the "acoustic" events, arise - apart from their differing SDDs and external shielding -  because Stage 1 ran in the Winter months: the "acoustic" events, which cover a wide range of background environmental noises, are larger since this season is generally more noisy. The radon levels in the LSBB complex varied between the Stages, being $\sim$ 50 Bq/m$^{3}$ and $\sim$ 3000 Bq/m$^{3}$ respectively, which were only partially reduced by the various suppression methods described above; since all Stage 2 SDDs were fabricated in the higher radon field, their gel contained more radon - as seen in terms of increased $\alpha$ events (0.95/day for Stage 2 vs. 0.69/day for Stage 1) and in resulting fractures. The Stage 2 decrease in higher $\mathcal{F}$ acoustics is associated with cable motion, resulting from improved cable securing during the shield reconstruction. The increased $\mathcal{A}$ of the acoustic events (specifically mechanical contacts) in Stage 2 is a result of the explosive failure of the four SDDs during the measurements.

In Stage 1, there were a total of 4048 signals recorded, of which 1820 were uncorrelated single events. The analysis identified 88\% with various environmental acoustic noise events, 3.4\% in trapped N$_{2}$ gas, 0.11\% in N$_{2}$ escape, and 4.4\% in fractures. Reanalysis of the Stage 1 recoil signals following Ref. \cite{prl1} identified 4 of the previously-identified 14 recoil events with exponential decay characteristic of nonuniform impulses observed in acoustic background studies associated with SDDs in vibrational contact with their support and air bubbles from water inflow, reducing the recoil events to 10; removal of the 3 recoil events recorded during the installation phase reduces this to 7. Five $\alpha$ events were also removed with the installation data.

In Stage 2, with its improved neutron shielding, there were a total of 1982 events of which 811 were uncorrelated single events, from which the analysis identified 83\% with various environmental noise events, 3.9\% in trapped N2 gas and 8.0\% in fractures; a single event within the "true nucleation" frequency window and $\mathcal{A} <$ 100 mV was recorded. Discounting the 3 detector failures, only 41 fracture events were recorded over the first 45 days of the 12 detector operation, or an average of 3.4 fractures per detector; $\sim$ 40\% of all recorded fractures occurred in the 3 SDDs which failed early in the Stage, and the fracture rate is otherwise roughly consistent with Stage 1. A synopsis of the particle-induced results for each SDD of the two Stages is shown in Table IV, with "P" the freon-less SDD monitor. The locations of the SDD's (A-P) in the water pool varied between the two Stages. Correlations between positions and rates of true nucleations were investigated; none were identified.

\begin{table*}
\caption{\label{table} Total particle-induced (P-i) events (evts)in each of the SDDs in the two Stages of the Phase II measurement with Stage 1 corrected for the inadvertent inclusion of its 1.94 kgd "installation" data; the first three SDDs failing in Stage 2 (I, L, M) are set to "0", since their data records were not used. The "P" detector in both cases was freon-less.}
\begin{ruledtabular}
\begin{tabular}{ccccccccccccccccc}
  position & A & B & C & D & E & F & G & H & I & J & K & L & M & N & O & P    \\
  \hline
  P-i: $\leq$ 2.2 bar (evts) & 6  & 4  & 5  & 6   & 0  & 0  & 0  & 16 & 0 & 3 & 0 & 0 & 0 & 2 & 6 & 0    \\
  \hline
  P-i: $\leq$ 2.2 bar (evts) & 0  & 0  & 0  & 0   & 2  & 0  & 5  & 7 & 0 & 0 & 2 & 0 & 0 & 3 & 11 & 0  \\
\end{tabular}
\end{ruledtabular}
\end{table*}

\subsection{Background Estimates}

The $\alpha$ contribution to the measurements was analyzed by time-dependent diffusion of the atmospheric radon concentration through the water and material components. The measured presence of U/Th contaminations in the gel yielded an overall $\alpha$-background level of $<$ 0.5 evt/kg freon/d. The overall $\alpha$ contribution to the Stage 1 measurement, including both the radon and U/Th decay progeny, was estimated at 3.26 $\pm$ 0.08 (stat) $\pm$ 0.76 (syst) evt/kgd, where $\sigma_{stat}$ derives from the simulation statistics and $\sigma_{syst}$ is obtained from the materials' characterizations and simulation geometry. The $\alpha$ contribution to the Stage 2 measurement was similarly 5.72 $\pm$ 0.12 (stat) $\pm$ 0.29 (syst) evt/kgd.

The individual neutron-induced recoil contribution estimates, shown in Table III, were estimated via MCNP simulations which included all materials' U/Th radio-assays for both detectors and shield. For Stage 1, the total expected background was 0.976 $\pm$ 0.004 (stat) $\pm$ 0.042 (syst) evt/kgd; for Stage 2, with the increased PE+wood+paraffin shielding, and improved shielding construct, the estimated background was 0.33 $\pm$ 0.001 (stat) $\pm$ 0.038 (syst) evt/kgd, with the principle contributions from detector glass (98\%) and unpurified shield water (0.8\%); without the glass, the overall rate is reduced to 10$^{-3}$ evt/kgd with the principle contribution from the concrete.

\section{Data Interpretations}

\subsection{Analyses}

Following completion of both the signal and background analyses, the two results were unblinded for comparison. The $\alpha$ yield over the two exposures was 41 and 29 events in Stage 1 and 2 respectively, in good agreement with their respective background estimates. The 7 recoil events in Stage 1 are below the exposure-corrected, estimated 10.8$\pm$0.5 background neutron-induced recoils; the 1 recoil event in Stage 2 was similarly below the anticipated 1.9$\pm$0.2 background events.

The agreement with the particle-induced simulation estimates in both Stages suggests the recoil events to be of neutron origin, and the results are treated as before with the adopted "97\% acceptance" for the recoil identification cut, and a conservative Feldman-Cousins (F-C) method \cite{feldman} in which the neutron-induced recoil background estimate in each Stage is reduced by $\sigma_{syst}$ to account for estimate uncertainties.

Given the threshold nature of the SDD, the resulting rates were analyzed in the standard framework by integrating

\begin{equation}\label{rate}
    \frac{dR}{dE_{R}} = \frac{\rho \epsilon^{A}}{2M_{W}}\frac{\sigma_{A}}{\mu_{A}^{2}}F^{2} \int_{v_{min}}^{v_{max}} \frac{f(v)}{v} d^{3}v
\end{equation}

\noindent over E$_{R}$, where $\rho$ is the local WIMP halo mass density, M$_{W}$ is the WIMP mass, $\mu_{A}$ is the reduced mass, $F^{2}(E_{R})$ is the Helm nuclear form factor, $\epsilon^{A}$ is the detector efficiency of Eq. (5), $f(v)$ is the halo model-dependent WIMP velocity distribution function, with v$_{max}$ = the galactic escape velocity (v$_{esc}$) + the earth's motion relative to the Galaxy (v$_{E}$), and the lower limit of the velocity integral given by

\begin{equation}\label{vmin}
    v_{min} = \sqrt{\frac{M_{W}E_{thr}^{nr}}{2\mu_{A}^{2}}}  .
\end{equation}

\noindent The zero-momentum transfer cross section $\sigma_{A}$ is the sum of spin-dependent (SD) and -independent (SI) contributions (recently shown to be incomplete \cite{fitz1,fitz2}), written as:

\begin{eqnarray}\label{xsection}
    \sigma_{A}^{SD} = 8\chi [(a_{p}^{A}<S_{p}^{A}>+a_{n}^{A}<S_{n}^{A}>)^{2}\frac{J+1}{J}]\\
    \sigma_{SI} = \chi [f_{p}^{A}Z+f_{n}^{A}N]^{2},
\end{eqnarray}

\noindent where $\chi = \frac{4}{\pi}G_{F}^{2}\mu_{A}^{2}$, G$_{F}$ is the Fermi constant, f$_{p,n}^{A}$ (a$_{p,n}^{A}$) are the SI (SD) WIMP couplings with the proton/neutron respectively,  $<$S$_{p,n}^{A}>$ are the ensemble target nucleus proton and neutron spins, and J is the total nuclear spin of each target constituent.

The analysis employed the Lewin-Smith (L-S) \cite{lewin} parametrization of $F$, as well as the standard isotropic isothermal halo model (SHM) parameters without the correction to the distribution near the cutoff velocity \cite{savage} as a means of normalizing comparisons between various experimental results: $\rho$ = 0.3 GeV/cm$^{3}$, v$_{0}$ = 230 km/s, v$_{esc}$ = 600 km/s and v$_{E}$ = 244 km/s.

For the SI sector, the isospin-conserving $f_{p}^{A}$ = $f_{n}^{A}$ was assumed. The two spin sectors were evaluated simultaneously (a$_{p,n} \neq$ 0), using the values of Pacheco-Strottman (P-S) ($<S_{p}^{F}>$ = 0.441, $<S_{n}^{F}>$ = -0.109) for $^{19}$F \cite{strottman}; for $^{35}$Cl and $^{37}$Cl, the $<S_{p,n}^{Cl}>$ were from Ref. \cite{mi}, while the $<S_{p,n}^{C}>$ were estimated for $^{13}$C by using the odd group approximation. The SD $\sigma_{p,n}$ contours were computed following Ref. \cite{mi} in which the rate is distributed over all target nuclei of the C$_{2}$ClF$_{5}$. Both SD and SI contours included 1$\sigma$ uncertainties in $\Lambda$ and $\Gamma$ which yielded variations in the results of $\leq$ 1\%.

The resulting contours are presented in Figs. 17 - 19 ("SIMPLE-2014"), respectively, with 4.3 $\times$ 10$^{-3}$ pb at 35 Gev/c$^{2}$ (S$_{p}$D) and 3.6 $\times$ 10$^{-6}$ pb at 35 Gev/c$^{2}$ (SI); these are shown in comparison with the previous report ("SIMPLE-2012") of Ref. \cite{prl2}. Fig. 18, displaying the resulting S$_{n}$D contour with minimum of 7.1 $\times$ 10$^{-2}$ pb at 30 Gev/c$^{2}$, is included for completeness following Ref. \cite{sdn}.

\subsection{Re-examinations}

As stated above, various aspects of the results' interpretations were questioned following Ref. \cite{prl2}, which are discussed below.

\subsubsection{Recoil Detection Rate}

Application of the F-C method, with the neutron-induced recoil background estimate in each Stage reduced by $\sigma_{syst}$, was questioned as being overly conservative. The strict F-C method does not include any uncertainty in the background estimate, but rather assumes the background is perfectly known; without the 1$\sigma$ reduction, the reported contour minima in the S$_{p}$D and SI sectors are lower by 6\% respectively, as shown in Figs. 17 and 19 ("F-C only"). The true limit should lie somewhere between these two pairs of contours.

\subsubsection{Fluorine spin}

The use of the P-S $<S_{p,n}^{F}>$ was also questioned \cite{cannoni}, given the work of Divari et. al. ($<S_{p}^{F}>$ = 0.475, $<S_{n}^{F}>$ = -0.009) \cite{divari} which in principle uses the more realistic interaction of Wildenthal. In Figs. 17 - 18, we show the impact of replacing the P-S $<$S$_{p,n}^{F}>$ with the Divari $<$S$_{p,n}^{F}>$ ("Divari only") in the analysis (all else unchanged): as seen, the reach of the "2014" contour in the S$_{p}$D sector is increased to 3.71$\times$10$^{-3}$ pb at the contour minimum, while it decreases to 1.03$\times$ 10$^{1}$ pb in the S$_{n}$D as anticipated. Note however that the other fluorine-based experiments also use the P-S $<$S$_{p,n}^{F}>$, so that similar changes would be obtained for each.

\begin{figure}[h]
  \includegraphics[width=9 cm]{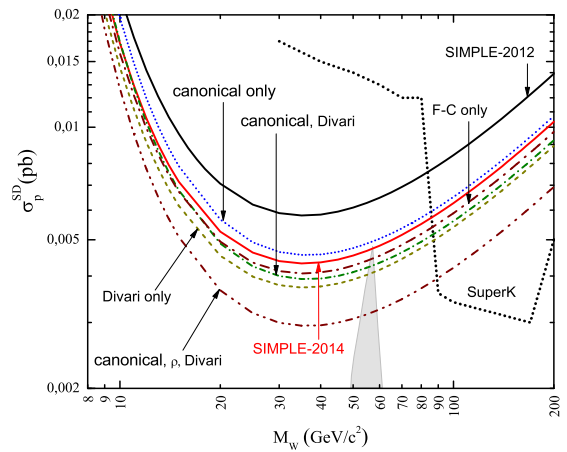}\\
  \caption{New S$_{p}$D exclusion contour ("SIMPLE-2014") resulting from correction of the Stage 1 results, together with its variations from the use of the strict Feldman-Cousins application ("F-C only"), Divari fluorine spin values ("Divari"), the canonical SHM v$_{0}$,v$_{esc}$ parameters ("canonical"), and canonical + $\rho_{0}$ = 0.40 GeV/c$^{2}$ ("canonical, $\rho$"); the previous "merged" contour, reported in Ref. \cite{prl2} using the Pacheco-Strottman $<$S$_{p,n}^{F}>$ \cite{strottman} and Lewin-Smith SHM parameters, is also shown ("SIMPLE-2012") for comparison. The region in gray denotes the area suggested by CMSSM \cite{cmssm}.}
  \label{fig3}
\end{figure}

\begin{figure}[h]
  \includegraphics[width=9 cm]{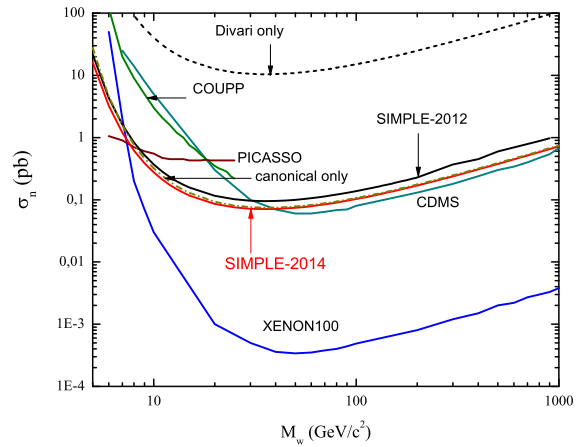}\\
  \caption{New $S_{n}$D exclusion contour ("SIMPLE-2014") resulting from the corrected Stage 1 results, together with that of the 2012 report ("SIMPLE 2012"). Also shown are contours for the "Divari only" and "canonical only".}
  \label{fig4}
\end{figure}

\subsubsection{Halo model parameters}

The parameters of the L-S SHM differ from those of v$_{0}$ = 220 km/s, v$_{esc}$ = 544 km/s in "canonical" use by other experiments \cite{green,reid, savage2}. In fact, while only PICASSO also uses the strict L-S parameters, only XENON \cite{xe100} and CRESST-II \cite{cresst2} use the canonical; most others (and several theoretical interpretations) have used somewhat different, "mix \& match" SHM parameter sets as indicated in Table V. All but CDMS and EDELWEISS use v$_{E}$ = 244 km/s, making it also "canonical". This however represents an average of a yearly cycle \cite{lewin}, and the variations in Table V likely reflect the execution calendar of the experiments; in the case of SIMPLE Stage 1 for example, $<$v$_{E} >$ = 231.9 km/s while for Stage 2, $<$v$_{E} >$ = 255.7 km/s (not however used in analysis).

The impact of using the canonical SHM parameters in the Phase II analysis is also shown ("canonical only") in Figs. 17 - 19, yielding $\sim$ 5\% weaker limits in each case although there is little change in the SI sector for M$_{W} >$ 50 GeV/c$^{2}$. As discussed in Refs. \cite{mccabe,green2}, a smaller v$_{0}$ in general shifts the exclusion contours to larger $\sigma_{A}$ and M$_{W}$, as is also the case with a smaller v$_{esc}$ for M$_{W} \leq$ 10 GeV/c$^{2}$. Generally, shifts due to changes in v$_{0}$ are larger than from similar changes in v$_{esc}$. Both parameters relate to the SHM distribution sampled by the experiment, which further depends on E$_{thr}^{nr}$ and v$_{min}$ of the detector. As seen in the S$_{p}$D sector, the new "canonical" contour with the Divari $<$S$_{p,n}^{F}>$ reaches to 3.91$\times$10$^{-3}$ pb at 35 GeV/c$^{2}$. Note however that more recent analysis has suggested v$_{0}$ as large as 280 km/s \cite{green,reid,gondolo}, which would improve the reach of all experiments as well as their low M$_{W}$ sensitivity.

In the SI sector, use of the CoGeNT-2008 SHM parameters with v$_{E}$ = 650 km/s and v$_{0}$ = v$_{0}^{L-S}$ yields virtually no change in the SIMPLE SI contour, whereas use of the EDELWEISS-2011 set with its v$_{0}$ = 270 km/s and decreased v$_{E}$ yields $\sim$ 13\% reduction with the contour minimum shifted to $\sim$ 35 GeV/c$^{2}$; at 10 GeV/c$^{2}$, the SIMPLE contour is less restrictive by $\sim$ 36\%.

\begin{table}[h]
\caption{\label{table}Survey of Standard Halo Model parameters in recent use by leading experiment reportings.}
\begin{ruledtabular}
\begin{tabular}{ccccccccccccccccc}
  experiment & v$_{0}$  & v$_{esc}$  & v$_{E}$ & $\rho_{0}$  \\
  & (km/s) & (km/s) & (km/s) & (GeV/cm$^{3}$) \\
  \hline
   L \& S \cite{lewin} & 230 & 600 & 244 & 0.3 \\
   canonical \cite{green} & 220 & 544 & 244 & 0.3 \\
   \hline
   COUPP \cite{coupp2012} & 230 & 544 & 244 & 0.3 \\
   PICASSO \cite{pic2012} & 230 & 600  & 244  & 0.3\\
   KIMS-2012 \cite{kims2012} & 220 & 650  & 244  & 0.3\\
   CoGeNT-2008 \cite{cogent08} & 230 & 650 & 244 & 0.3 \\
   CoGeNT-2012 \cite{cogent12} & 220 & 550 & 244 & 0.3 \\
   CDMS/EDELWEISS \cite{cdmsedel} & 220 & 544 & 232 & 0.3 \\
   CDMS-Ge \cite{cdms2011} & 220 & 544 & 232 & 0.3 \\
   CDMS-Si \cite{cdms2013} & 220 & 544 & 232 & 0.3 \\
   EDELWEISS-2011 \cite{edel2011} & 270 & 544 & 235 & 0.3 \\
   EDELWEISS-lite \cite{edelite} & 270 & 544 & 235 & 0.3 \\
   CRESST-II \cite{cresst2} & 220 & 544 & 244 & 0.3 \\
   ZEPLIN \cite{zeplin} & 220  & 544 & 232 & 0.3 \\
   XENON10 \cite{xenon10} & 230  & 544 & 244 & 0.3 \\
   XENON100 \cite{xe100} & 220  & 544 & 244 & 0.3 \\
   LUX \cite{lux} & 220  & 544 & 245 & 0.3 \\
\end{tabular}
\end{ruledtabular}
\end{table}

\begin{figure}[h]
  \includegraphics[width=10 cm]{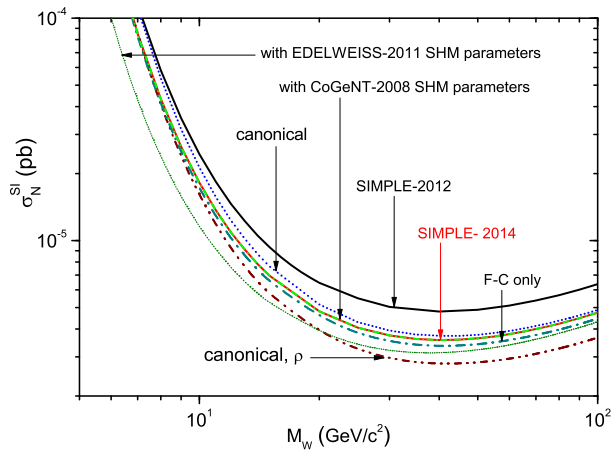}\\
  \caption{New SI exclusion contour ("SIMPLE-2014") resulting from correction of the Stage 1 results, together with its variations  using a strict Feldman-Cousins application ("FC only"), the canonical SHM parameters ("canonical"), and canonical+$\rho$=0.40 GeV/c$^{2}$ ("canonical, $\rho$"); the previous "merged" contour, reported in Ref. \cite{prl2} using Lewin-Smith SHM parameters, is also shown ("SIMPLE-2012") for comparison}.
  \label{fig3}
\end{figure}

Although $\rho$ = 0.3 GeV/cm$^{3}$ is used in all reports, this parameter has also come under recent scrutiny, with a survey indicating $\rho_{0}$ = 0.20 - 0.55 GeV/cm$^{3}$ \cite{iocco}, and $\rho$ $\sim$ 0.40 more favored \cite{salucci,catena,freese}. Since dE/dE$_{R}$ is directly proportional to $\rho$, larger values would reduce any cross section derived from the data independent of the SHM velocity-structure (note however that v$_{0}$ and $\rho$ are related - see Ref. \cite{mccabe}). The impact of $\rho$ = 0.40 GeV/cm$^{3}$ together with the canonical velocities is also shown ("canonical, $\rho$") in Figs. 17 - 19, demonstrating the improved reach which would be achieved by all experimental results. Recent evidence further suggests the halo distribution to differ from the isotropic SHM, resulting in parameters differing from those considered herein. Examination of the impact of these is however well beyond the scope of this report.

\section{SUMMARY}

We have provided a detailed overview of the SIMPLE Phase II measurements, to include a correction of the Stage 1 data for the inadvertent inclusion of the detector installation data into the previous science report. Reanalysis of the overall result provides $\sim$ 25\% improved limit contours in both the spin-dependent and -independent sectors. The overall number of recorded "true nucleation" events reflects the general insensitivity of the SDD to a majority of the common search experiment backgrounds, as well as an ability to significantly reduce neutron-generated recoil backgrounds to below the current level solely by improvement in the SDD containment shield. The differential agreement between the measurements and estimates of the two Stages supports the quality of the background estimates, the recoil nature of the recorded events, and their neutron origin.

The impact of any result depends foremost on the definition of the nuclear recoil acceptance window, in which SIMPLE differs from PICASSO and COUPP. For SIMPLE, the $\alpha$-descrimination is based on a well-resolved gap in the signal ln($\mathcal{A}^{2})$ between the two distributions, which derives from their respective LETs, and matching of the $\alpha$ dE/dx Bragg peak with the mean of the distribution of droplet sizes to provide a natural lower cutoff to the recorded $\alpha$ energy deposition; the cutoff is basically responsible both for the observed amplitude gap between the $\alpha$ and neutron recoil populations and the spectral asymmetry in the $\alpha$ distribution. Although the description neglects $\alpha$ event origins outside the droplet surfaces (with the complication of the $\alpha$ dE/dx in the gel being similar to that in C$_{2}$ClF$_{5}$) and requires more detailed examination, it nonetheless captures the essence of the involved physics. The reasons for the difference with PICASSO and COUPP -- while still not completely understood -- can at least be partially attributed to the difference in LET$_{c}$ of the different superheated liquids at their respective operating temperatures/pressures.

The results of both the current and previous \cite{prl2} analysis continue to be somewhat conservative via use of (i) a "modified" Feldman-Cousins analysis, (ii) the Pacheco-Strottman fluorine spin, and (iii) the Lewin-Smith SHM parameters in the analysis. A singular replacement of the Pacheco-Strottman $<$S$_{p,n}^{F}>$ by the Divari $<$S$_{p,n}^{F}>$ yields an improved contour reach in the SD sector (as also the PICASSO and COUPP results).

As evident, the impact of any measurement is strongly dependent on the halo model employed. Each of the SHM parameters is stated with $\sim$ 10\% uncertainty, which is not negligible in the interpretations since the dispersion between their measurements is larger \cite{green2}. The re-establishment of a common "standard model/parameter set" as a basis for analysis of all experimental results would obviate these confusions in all experiment comparisons, as well as simplify their use in theoretical interpretations. A complementary, halo model-independent approach based on the separability of the velocity integral of Eq. (11), has recently been suggested \cite{weiner}: although the SIMPLE result has been recast in this formalism \cite{frandsen,gondolo}, these consider only the Stage 2 result, and will be considered in a forthcoming paper, The approach is itself however complicated by the necessity of choosing  M$_{W}$ for presentation of the results (not unlike as with the model-independent representation \cite{mi} in terms of a$_{p,n}$) hence itself requiring a standardization.


With the conclusion of Phase II, SIMPLE however ends its use of SDDs in in the WIMP search effort in consequence of their low active ingredient-to-detector mass ratio. In its duration, Phase II has demonstrated the viability of high concentration, relatively inexpensive C$_{2}$ClF$_{5}$ SDD constructions in the search for astroparticle dark matter evidence, to include the chemistry necessary to achieve detector fabrications with a demonstrated stable operation over extended periods at operating temperatures and pressures capable of providing low energy recoil threshold measurements, providing results comparable with COUPP via $\sim$ 24x less exposure and superior to PICASSO, with $\sim$ 6x less. It has further demonstrated the  ability to construct similar SDDs with a variety of superheated liquids towards eventually providing more restrictive results via different target comparisons, and the capacity of a refined low frequency instrumentation to provide discrimination of nuclear recoil events from all acoustic gel-associated, environmental, and $\alpha$-induced backgrounds. Many of these properties have application in other areas of particle-detection physics, including neutron dosimetry and spectroscopy, $\alpha$-dosimetry and heavy ion reactions, and are currently being explored.

For dark matter applications, an essentially completed SIMPLE Phase III comprises a transition to larger mass, higher concentration bubble chamber technology using its gel/glycerin-sheathed containment to reduce spontaneous nucleation events, and many of the lessons gained from the previous phase. Several prototype chambers, each $\sim$ 4x the Phase II active mass, have been developed and are currently undergoing testing prior a rescaling to 20 kg devices and the construction of a Phase IV modular 1 ton detector.

\begin{acknowledgments}

We foremost thank the LSBB management and staff for their continual support and assistance over these years, and Raul Martins for his many suggestions and guidance in the instrumentation transition from Phase 1 to Phase 2. We have benefitted from the critical comments of Franco Giuliani during the course of this work, as well as invaluable radioassays of the site concrete and steel by Pia Loaiza. We further thank the Referees of Ref. \cite{prl2} and this work for their persistent, numerous criticisms and suggestions on the recoil window definitions, which forced a a significant review our thinking in making our replies. Discussions with Paolo Gondolo, Leo Stodolsky and Andrzej Drukier have been appreciated. We thank Jos\'e Albuquerque of CRIOLAB, Lda for numerous construction assistances during the experiment staging, Ant\'onio da Silva of ACP for his numerous transport arrangements, Ana Gouveia for her assistance in the Stage 1 radioassays of the shielding materials, and Georges Waysand and Denis Limagne for their continuing advices and interventions over the last many years (in particular the construction of the LSBB "white room" and transfer of the Am/Be source to the LSBB). We also sincerely thank the 12 students of Stephane Gaffet's 2010 geophysics class for their assistance in the not-so-simple Stage 2 shielding reconstruction, the gracious hospitality of the \textit{Patria}, and the Escoffier's, Casoli's, \textit{L'Aptois} and Mairie of Rustrel for each of their many kindnesses and support during our various residences near the LSBB over the Phase II years. We lastly thank Frank Avignone for several significant discussions and for his input in 2011.

This work was supported by grants PDTC/FIS/83424/2006, PDTC/FIS/115733/2009 and PDTC/FIS/121130/2010 of the Portuguese Foundation for Science and Technology (FCT), and the Nuclear Physics Center of the University of Lisbon. The activity of M. Felizardo was supported by grant SFRH/BD/46545/2008 of FCT.

\end{acknowledgments}


\begin{thebibliography}{21} \expandafter\ifx\csname
natexlab\endcsname\relax\def\natexlab#1{#1}\fi
\expandafter\ifx\csname bibnamefont\endcsname\relax
  \def\bibnamefont#1{#1}\fi
\expandafter\ifx\csname bibfnamefont\endcsname\relax
  \def\bibfnamefont#1{#1}\fi
\expandafter\ifx\csname citenamefont\endcsname\relax
  \def\citenamefont#1{#1}\fi
\expandafter\ifx\csname url\endcsname\relax
  \def\url#1{\texttt{#1}}\fi
\expandafter\ifx\csname urlprefix\endcsname\relax\def\urlprefix{URL
}\fi \providecommand{\bibinfo}[2]{#2}
\providecommand{\eprint}[2][]{\url{#2}}

\bibitem[{\citenamefont{Felizardo}(2012)}]{prl2}
\bibinfo{author}{\bibfnamefont{M.} \bibnamefont{Felizardo}}, \bibnamefont{et al.}:
\bibinfo{journal}{Phys. Rev. Lett.}
\textbf{\bibinfo{volume}{108}}, \bibinfo{pages}{201302}
(\bibinfo{year}{2012}).

\bibitem[{\citenamefont{Behnke}(2012)}]{coupp2012}
\bibinfo{author}{\bibfnamefont{E.} \bibnamefont{Behnke}}, \bibnamefont{et al.}:
\bibinfo{journal}{Phys. Rev.} \textbf{\bibinfo{volume}{D86}},
\bibinfo{pages}{052001} (\bibinfo{year}{2012}).

\bibitem[{\citenamefont{Archambault}(2012)}]{pic2012}
\bibinfo{author}{\bibfnamefont{S.} \bibnamefont{Archambault}}, \bibnamefont{et al.}:
\bibinfo{journal}{Phys. Lett.} \textbf{\bibinfo{volume}{B711}},
\bibinfo{pages}{153} (\bibinfo{year}{2012}).

\bibitem[{\citenamefont{Seitz}(1958)}]{seitz}
\bibinfo{author}{\bibfnamefont{F.} \bibnamefont{Seitz}}:
\bibinfo{journal}{Phys. Fluids} \textbf{\bibinfo{volume}{1}},
\bibinfo{pages}{2} (\bibinfo{year}{1958}).

\bibitem[{\citenamefont{lsbb}(2012)}]{lsbb}
\bibinfo{journal}{LSBB, http://lsbb.oca.eu}
(\bibinfo{year}{2012}).

\bibitem[{\citenamefont{Felizardo}(2010)}]{prl1}
\bibinfo{author}{\bibfnamefont{M.} \bibnamefont{Felizardo}} \bibnamefont{et al.}:
\bibinfo{journal}{Phys. Rev. Lett.} \textbf{\bibinfo{volume}{105}},
\bibinfo{pages}{211301} (\bibinfo{year}{2010}).

\bibitem[{\citenamefont{Collar}(2011)}]{cmt1}
\bibinfo{author}{\bibfnamefont{J.I.} \bibnamefont{Collar}}:
\bibinfo{journal}{arxiv:1106.3559v1}
(\bibinfo{year}{2011}).

\bibitem[{\citenamefont{Dahl}(2012)}]{cmt2}
\bibinfo{author}{\bibfnamefont{C. E.} \bibnamefont{Dahl}},
\bibinfo{author}{\bibfnamefont{J.} \bibnamefont{Hall}},
\bibinfo{author}{\bibfnamefont{H.} \bibnamefont{Lippincot}}:
\bibinfo{journal}{Phys. Rev. Lett.} \textbf{\bibinfo{volume}{108}},
\bibinfo{pages}{259001} (\bibinfo{year}{2012}).

\bibitem[{\citenamefont{Girard}(2011)}]{reply1}
\bibinfo{author}{\bibfnamefont{TA} \bibnamefont{Girard}}:
\bibinfo{journal}{arxiv:1106.3559v1}
(\bibinfo{year}{2011}).

\bibitem[{\citenamefont{Girard}(2012)}]{reply2}
\bibinfo{author}{\bibfnamefont{TA} \bibnamefont{Girard}}, \bibnamefont{et al.}:
\bibinfo{journal}{Phys. Rev. Lett.} \textbf{\bibinfo{volume}{108}},
\bibinfo{pages}{259002} (\bibinfo{year}{2012}).

\bibitem[{\citenamefont{Puibasset}(2000)}]{jptese}
\bibinfo{author}{\bibfnamefont{J.} \bibnamefont{Puibasset}}:
\bibinfo{journal}{PhD thesis, Univ. Paris 7, 2000 (unpublished)}.

\bibitem[{\citenamefont{Collar}(2000)}]{njp}
\bibinfo{author}{\bibfnamefont{J.I.} \bibnamefont{Collar}}, \bibnamefont{et al.}:
\bibinfo{journal}{New J. Phys} \textbf{\bibinfo{volume}{2}},
\bibinfo{pages}{14.1} (\bibinfo{year}{2000}).

\bibitem[{\citenamefont{Morlat}(2000)}]{tmtese}
\bibinfo{author}{\bibfnamefont{T.A} \bibnamefont{Morlat}}:
\bibinfo{journal}{PhD thesis, Univ. Paris 7, 2004 (unpublished)}.

\bibitem[{\citenamefont{Morlat}(2006)}]{tomonim}
\bibinfo{author}{\bibfnamefont{T.}~\bibnamefont{Morlat}}
\bibnamefont{et al.}:
\bibinfo{journal}{Nucl. Instr. \& Meth.}
\textbf{\bibinfo{volume}{A560}},
\bibinfo{pages}{339} (\bibinfo{year}{2006}).

\bibitem[{\citenamefont{Felizardo}(2009)}]{felizimprov}
\bibinfo{author}{\bibfnamefont{M.} \bibnamefont{Felizardo}},
\bibnamefont{et al.}:
\bibinfo{journal}{Nucl. Instrum. \& Meth.} \textbf{\bibinfo{volume}{A585}},
\bibinfo{pages}{61} (\bibinfo{year}{2008}).

\bibitem[{\citenamefont{Felizardo}(2009)}]{feliznew}
\bibinfo{author}{\bibfnamefont{M.} \bibnamefont{Felizardo}},
\bibnamefont{et al.}:
\bibinfo{journal}{Nucl. Instrum. \& Meth.} \textbf{\bibinfo{volume}{A589}},
\bibinfo{pages}{72} (\bibinfo{year}{2008}).

\bibitem[{\citenamefont{Felizardo}(2013)}]{mftese}
\bibinfo{author}{\bibfnamefont{M.} \bibnamefont{Felizardo}}:
\bibinfo{journal}{PhD thesis, Univ. Nova de Lisboa, 2013 (unpublished)}.

\bibitem[{\citenamefont{Ghilea}(2011)}]{ghilea}
\bibinfo{author}{\bibfnamefont{M.C.} \bibnamefont{Ghilea}} \bibnamefont{et al.}:
\bibinfo{journal}{Nucl. Instrum. \& Meth.}
\textbf{\bibinfo{volume}{A648}},
\bibinfo{pages}{210}(\bibinfo{year}{2011}).

\bibitem[{\citenamefont{d'Errico}(2001)}]{derrico}
\bibinfo{author}{\bibfnamefont{F.}~\bibnamefont{d'Errico}}:
\bibinfo{journal}{Nucl. Instrum. \& Meth.}
\textbf{\bibinfo{volume}{B184}},
\bibinfo{pages}{229}(\bibinfo{year}{2001}).

\bibitem[{\citenamefont{Sun}(1992)}]{sun}
\bibinfo{author}{\bibfnamefont{Y.Y.}~\bibnamefont{Sun}},
\bibinfo{author}{\bibfnamefont{B.T.}~\bibnamefont{Chu}},
\bibinfo{author}{\bibfnamefont{R.E.}~\bibnamefont{Apfel}}:
\bibinfo{journal}{Journ. Comput. Phys.}
\textbf{\bibinfo{volume}{103}},
\bibinfo{pages}{116}(\bibinfo{year}{1992}).

\bibitem[{\citenamefont{Martynyuk}(2008)}]{marty}
\bibinfo{author}{\bibfnamefont{Yu.N.}~\bibnamefont{Martynyuk}},
\bibinfo{author}{\bibfnamefont{N.S.}~\bibnamefont{Smirnov}}:
\bibinfo{journal}{Sov. Phys. Acoust.}
\textbf{\bibinfo{volume}{37}},
\bibinfo{pages}{376}(\bibinfo{year}{1991}).

\bibitem[{\citenamefont{Plesset}(1954)}]{plesset}
\bibinfo{author}{\bibfnamefont{M.S}~\bibnamefont{Plesset}},
\bibinfo{author}{\bibfnamefont{S.A} \bibnamefont{Zwick}}:
\bibinfo{journal}{J. Appl. Phys.}
\textbf{\bibinfo{volume}{25}},
\bibinfo{pages}{493} (\bibinfo{year}{1954}).

\bibitem[{\citenamefont{Harper}(1991)}]{harper}
\bibinfo{author}{\bibfnamefont{M.} \bibnamefont{Harper}}:
\bibinfo{journal}{PhD thesis, Univ. Maryland, 1991 (unpublished)}.

\bibitem[{\citenamefont{Duan}(2000)}]{duan}
\bibinfo{author}{\bibfnamefont{Y.Y.} \bibnamefont{Duan}},
\bibnamefont{et al.}:
\bibinfo{journal}{Int. J. Thermophys.} \textbf{\bibinfo{volume}{21}},
\bibinfo{pages}{393} (\bibinfo{year}{2000}).

\bibitem[{\citenamefont{NIST}(2002)}]{nist}
\bibinfo{author}{NIST, REFPROP code},
\bibinfo{journal}{NIST Standard Database} \textbf{\bibinfo{volume}{23}},
\bibinfo{pages}{V.7} (\bibinfo{year}{2002}).

\bibitem[{\citenamefont{Bonin}(2001)}]{bonin}
\bibinfo{author}{\bibfnamefont{H.W.} \bibnamefont{Bonin}},
\bibnamefont{et al.}:
\bibinfo{journal}{Rad. Prot. Dosim.} \textbf{\bibinfo{volume}{46}},
\bibinfo{pages}{265} (\bibinfo{year}{2001}).

\bibitem[{\citenamefont{Barnab\'e-Heider}(2005)}]{barnabe}
\bibinfo{author}{\bibfnamefont{M.}~\bibnamefont{Barnab\'e}},
\bibnamefont{et al.}:
\bibinfo{journal}{Nucl. Instr. \& Meth.}
\textbf{\bibinfo{volume}{A555}},
\bibinfo{pages}{184} (\bibinfo{year}{2005}).

\bibitem[{\citenamefont{Girard}(2011)}]{srim}
\bibinfo{journal}{http://www.srim.org/}

\bibitem[{\citenamefont{Eberhard}(1975)}]{eber}
\bibinfo{author}{\bibfnamefont{J.G.}~\bibnamefont{Eberhard}},
\bibnamefont{et al.}:
\bibinfo{journal}{J. Colloid Interf. Sci.}
\textbf{\bibinfo{volume}{56}},
\bibinfo{pages}{369} (\bibinfo{year}{1975}).

\bibitem[{\citenamefont{Landau}(1987)}]{landau}
\bibinfo{author}{\bibfnamefont{L.D.}~\bibnamefont{Landau}},
\bibinfo{author}{\bibfnamefont{E.M.}~\bibnamefont{Lifschitz}}:
\bibinfo{journal}{Fluid Mechanics, Vol. VI (Pergamon, UK, 1987)}.

\bibitem[{\citenamefont{Felizardo}(2013)}]{felizfabvar}
\bibinfo{author}{\bibfnamefont{M.}~\bibnamefont{Felizardo}},
\bibinfo{author}{\bibfnamefont{T.} \bibnamefont{Morlat}},
\bibinfo{author}{\bibfnamefont{J.G.} \bibnamefont{Marques}}, \bibnamefont{et al.}:
\bibinfo{journal}{Astrop. Phys.}
\textbf{\bibinfo{volume}{49}},
\bibinfo{pages}{28} (\bibinfo{year}{2013}).

\bibitem[{\citenamefont{Minnaert}(1933)}]{minnaert}
\bibinfo{author}{\bibfnamefont{M.}~\bibnamefont{Minnaert}}:
\bibinfo{journal}{Phil. Mag.}
\textbf{\bibinfo{volume}{16}},
\bibinfo{pages}{235} (\bibinfo{year}{1933}).

\bibitem[{\citenamefont{Giuliani}(2004)}]{itn}
\bibinfo{author}{\bibfnamefont{F.} \bibnamefont{Giuliani}}, \bibnamefont{et al.}:
\bibinfo{journal}{Nucl. Instr. \& Meth.} \textbf{\bibinfo{volume}{A526}},
\bibinfo{pages}{526} (\bibinfo{year}{2004}).

\bibitem[{\citenamefont{Morlat}(2008)}]{tomoap}
\bibinfo{author}{\bibfnamefont{T.A.}~\bibnamefont{Morlat}}, \bibnamefont{et.~al.}
\bibinfo{journal}{Astrop. Phys.}
\textbf{\bibinfo{volume}{30}},
\bibinfo{pages}{159}(\bibinfo{year}{2008}).

\bibitem[{\citenamefont{Felizardo}(2009)}]{felizspatloc}
\bibinfo{author}{\bibfnamefont{M.} \bibnamefont{Felizardo}},
\bibnamefont{et al.}:
\bibinfo{journal}{Nucl. Instrum. \& Meth.} \textbf{\bibinfo{volume}{A599}},
\bibinfo{pages}{93} (\bibinfo{year}{2009}).

\bibitem[{\citenamefont{Archambault}(2011)}]{newinsite}
\bibinfo{author}{\bibfnamefont{S.} \bibnamefont{Archambault}}, \bibnamefont{et al.}:
\bibinfo{journal}{New Journ. Phys.} \textbf{\bibinfo{volume}{13}},
\bibinfo{pages}{043006} (\bibinfo{year}{2011}).

\bibitem[{\citenamefont{Felizardo}(2010)}]{felizvary}
\bibinfo{author}{\bibfnamefont{M.} \bibnamefont{Felizardo}},
\bibnamefont{et al.}:
\bibinfo{journal}{Nucl. Instrum. \& Meth.} \textbf{\bibinfo{volume}{A614}},
\bibinfo{pages}{278} (\bibinfo{year}{2010}).

\bibitem[{\citenamefont{Girard}(2011)}]{inprog}
\bibinfo{journal}{M. Felizardo, et al (in preparation).}

\bibitem[{\citenamefont{Waysand}(2000)}]{waysand}
\bibinfo{author}{\bibfnamefont{G.}~\bibnamefont{Waysand}}, \bibnamefont{et al.}:
\bibinfo{journal}{Nucl. Instrum. \& Meth.}
\textbf{\bibinfo{volume}{A444}},
\bibinfo{pages}{336}(\bibinfo{year}{2000}).

\bibitem[{\citenamefont{amar\'e}(2006)}]{canfranc}
\bibinfo{author}{\bibfnamefont{J.}~\bibnamefont{Amar\'e}}, \bibnamefont{et al.}:
\bibinfo{journal}{Phys. Conf. Ser.}
\textbf{\bibinfo{volume}{39}},
\bibinfo{pages}{151}(\bibinfo{year}{2006}).

\bibitem[{\citenamefont{chazal}(1998)}]{modane}
\bibinfo{author}{\bibfnamefont{V.}~\bibnamefont{Chazal}}, \bibnamefont{et al.}:
\bibinfo{journal}{Astropart. Phys.}
\textbf{\bibinfo{volume}{9}},
\bibinfo{pages}{163}(\bibinfo{year}{1998}).

\bibitem[{\citenamefont{wulandari}(2004)}]{gransasso}
\bibinfo{author}{\bibfnamefont{H.}~\bibnamefont{Wulandari}}, \bibnamefont{et al.}:
\bibinfo{journal}{Astropart. Phys.}
\textbf{\bibinfo{volume}{22}},
\bibinfo{pages}{313}(\bibinfo{year}{2004}).

\bibitem[{\citenamefont{Fernandes}(2010)}]{acf}
\bibinfo{author}{\bibfnamefont{A.C.}~\bibnamefont{Fernandes}},
\bibnamefont{et al.}:
\bibinfo{journal}{Nucl. Instr. \& Meth.}
\textbf{\bibinfo{volume}{A623}},
\bibinfo{pages}{960} (\bibinfo{year}{2010}).

\bibitem[{\citenamefont{Shores}(2001)}]{shores}
\bibinfo{author}{\bibfnamefont{E.F.}~\bibnamefont{Shores}},
\bibnamefont{et al.}:
\bibinfo{journal}{Nucl. Instr. \& Meth.}
\textbf{\bibinfo{volume}{B179}},
\bibinfo{pages}{78} (\bibinfo{year}{2001}).

\bibitem[{\citenamefont{Mei}(2009)}]{mei}
\bibinfo{author}{\bibfnamefont{D.M.}~\bibnamefont{Mei}},
\bibnamefont{et al.}:
\bibinfo{journal}{Nucl. Instr. \& Meth.}
\textbf{\bibinfo{volume}{A606}},
\bibinfo{pages}{651} (\bibinfo{year}{2009}).

\bibitem[{\citenamefont{Chadwick}(2006)}]{chadwick}
\bibinfo{author}{\bibfnamefont{M.B.}~\bibnamefont{Chadwick}},
\bibnamefont{et al.}:
\bibinfo{journal}{Nucl. Data Sheets}
\textbf{\bibinfo{volume}{107}},
\bibinfo{pages}{2931} (\bibinfo{year}{2006}).

\bibitem[{\citenamefont{Feldman}(1998)}]{feldman}
\bibinfo{author}{\bibfnamefont{G.J.}~\bibnamefont{Feldman}},
\bibinfo{author}{\bibfnamefont{R.D.} \bibnamefont{Cousins}}:
\bibinfo{journal}{Phys. Rev.}
\textbf{\bibinfo{volume}{D57}},
\bibinfo{pages}{3873} (\bibinfo{year}{1998}).

\bibitem[{\citenamefont{Fitzpatrick}(2013)}]{fitz1}
\bibinfo{author}{\bibfnamefont{A.L.}~\bibnamefont{Fitzpatrick}},
\bibinfo{author}{\bibfnamefont{W.} \bibnamefont{Haxton}},
\bibinfo{author}{\bibfnamefont{E.} \bibnamefont{Katzl}},
\bibnamefont{at al.}:
\bibinfo{journal}{JCAP}
\textbf{\bibinfo{volume}{1302}},
\bibinfo{pages}{004} (\bibinfo{year}{2013}).

\bibitem[{\citenamefont{Fitzpatrick}(2013)}]{fitz2}
\bibinfo{author}{\bibfnamefont{A.L.}~\bibnamefont{Fitzpatrick}},
\bibinfo{author}{\bibfnamefont{W.} \bibnamefont{Haxton}},
\bibinfo{author}{\bibfnamefont{E.} \bibnamefont{Katzl}}
\bibnamefont{at al.}:
\bibinfo{journal}{arXiv:1211.2818 [hep-ph]}
(\bibinfo{year}{2012}).

\bibitem[{\citenamefont{lewin}(1996)}]{lewin}
\bibinfo{author}{\bibfnamefont{J.D.}~\bibnamefont{Lewin}},
\bibinfo{author}{\bibfnamefont{P.F.} \bibnamefont{Smith}}:
\bibinfo{journal}{Astropart. Phys.}
\textbf{\bibinfo{volume}{6}},
\bibinfo{pages}{87}(\bibinfo{year}{1996}).

\bibitem[{\citenamefont{Savage}(2006)}]{savage}
\bibinfo{author}{\bibfnamefont{C.}~\bibnamefont{Savage}},
\bibinfo{author}{\bibfnamefont{K.} \bibnamefont{Freese}},
\bibinfo{author}{\bibfnamefont{P.} \bibnamefont{Gondolo}}:
\bibinfo{journal}{Phys. Rev.}
\textbf{\bibinfo{volume}{D47}},
\bibinfo{pages}{043531}(\bibinfo{year}{2006}).

\bibitem[{\citenamefont{Strottman}(1989)}]{strottman}
\bibinfo{author}{\bibfnamefont{A.F.}~\bibnamefont{Pacheco}},
\bibinfo{author}{\bibfnamefont{D.} \bibnamefont{Strottman}}:
\bibinfo{journal}{Phys. Rev.}
\textbf{\bibinfo{volume}{D40}},
\bibinfo{pages}{2131}(\bibinfo{year}{1089}).

\bibitem[{\citenamefont{Giuliani}(2005)}]{mi}
\bibinfo{author}{\bibfnamefont{F.}~\bibnamefont{Giuliani}},
\bibinfo{author}{\bibfnamefont{TA} \bibnamefont{Girard}}:
\bibinfo{journal}{Phys. Rev.}
\textbf{\bibinfo{volume}{D71}},
\bibinfo{pages}{123503}(\bibinfo{year}{2005}).

\bibitem[{\citenamefont{Cannoni}(2011)}]{cannoni}
\bibinfo{author}{\bibfnamefont{M.} \bibnamefont{Cannoni}}:
\bibinfo{journal}{Phys. Rev.}
\textbf{\bibinfo{volume}{D84}},
\bibinfo{pages}{095017}
(\bibinfo{year}{2011}).

\bibitem[{\citenamefont{Divari}(2000)}]{divari}
\bibinfo{author}{\bibfnamefont{P.C.}~\bibnamefont{Divari}}, \bibnamefont{et al.}:
\bibinfo{journal}{Phys. Rev.}
\textbf{\bibinfo{volume}{C61}},
\bibinfo{pages}{054612}(\bibinfo{year}{2000}).

\bibitem[{\citenamefont{Roszkowski}(2010)}]{cmssm}
\bibinfo{author}{\bibfnamefont{L.}~\bibnamefont{Roszkowski}}, \bibnamefont{et al.}:
\bibinfo{journal}{J. High Energy Phys.}
\textbf{\bibinfo{volume}{07}},
\bibinfo{pages}{075}(\bibinfo{year}{2007}).

\bibitem[{\citenamefont{lippincott}(2013)}]{sdn}
\bibinfo{author}{\bibfnamefont{M.R.}~\bibnamefont{Buckley}},
\bibinfo{author}{\bibfnamefont{W.H.} \bibnamefont{Lippincott}}:
\bibinfo{journal}{Phys. Rev.}
\textbf{\bibinfo{volume}{D88}},
\bibinfo{pages}{056003}(\bibinfo{year}{2013}).

\bibitem[{\citenamefont{Green}(2010)}]{green}
\bibinfo{author}{\bibfnamefont{A.M.} \bibnamefont{Green}}:
\bibinfo{journal}{JCAP}
\textbf{\bibinfo{volume}{1010}},
\bibinfo{pages}{034}
(\bibinfo{year}{2010}).

\bibitem[{\citenamefont{Reid}(2009)}]{reid}
\bibinfo{author}{\bibfnamefont{M.J.}~\bibnamefont{Reid}}, \bibnamefont{et al.}:
\bibinfo{journal}{Astrophys. J.}
\textbf{\bibinfo{volume}{700}},
\bibinfo{pages}{137}(\bibinfo{year}{2009}).

\bibitem[{\citenamefont{Savage}(2009)}]{savage2}
\bibinfo{author}{\bibfnamefont{C.}~\bibnamefont{Savage}},
\bibinfo{author}{\bibfnamefont{G.}~\bibnamefont{Gelmini}},
\bibinfo{author}{\bibfnamefont{P.}~\bibnamefont{Gondolo}},
\bibinfo{author}{\bibfnamefont{K.} \bibnamefont{Freese}}:
\bibinfo{journal}{JCAP}
\textbf{\bibinfo{volume}{0904}},
\bibinfo{pages}{010}(\bibinfo{year}{2009}).

\bibitem[{\citenamefont{Aprile}(2013)}]{xe100}
\bibinfo{author}{\bibfnamefont{E.} \bibnamefont{Aprile}}, \bibnamefont{et al.}:
\bibinfo{journal}{Phys. Rev. Lett.}
\textbf{\bibinfo{volume}{111}},
\bibinfo{pages}{021301}
(\bibinfo{year}{2013}).

\bibitem[{\citenamefont{Angloher}(2012)}]{cresst2}
\bibinfo{author}{\bibfnamefont{G.} \bibnamefont{Angloher}}, \bibnamefont{et al.}:
\bibinfo{journal}{Eur. Phys. J.}
\textbf{\bibinfo{volume}{C72}},
\bibinfo{pages}{1971}
(\bibinfo{year}{2012}).

\bibitem[{\citenamefont{Kim}(2012)}]{kims2012}
\bibinfo{author}{\bibfnamefont{S.C.} \bibnamefont{Kim}}, \bibnamefont{et al.}:
\bibinfo{journal}{Phys. Rev. Lett.}
\textbf{\bibinfo{volume}{108}},
\bibinfo{pages}{181301}
(\bibinfo{year}{2012}).

\bibitem[{\citenamefont{Aalseth}(2008)}]{cogent08}
\bibinfo{author}{\bibfnamefont{C.E.} \bibnamefont{Aalseth}}, \bibnamefont{et al.}:
\bibinfo{journal}{Phys. Rev. Lett.}
\textbf{\bibinfo{volume}{101}},
\bibinfo{pages}{251301}
(\bibinfo{year}{2008}).

\bibitem[{\citenamefont{Aalseth}(2012)}]{cogent12}
\bibinfo{author}{\bibfnamefont{C.E.} \bibnamefont{Aalseth}}, \bibnamefont{et al.}:
\bibinfo{journal}{Phys. Rev.}
\textbf{\bibinfo{volume}{D88}},
\bibinfo{pages}{012002}
(\bibinfo{year}{2013}).

\bibitem[{\citenamefont{ahmed}(2011)}]{cdmsedel}
\bibinfo{author}{\bibfnamefont{Z.} \bibnamefont{Ahmed}}, \bibnamefont{et al.}:
\bibinfo{journal}{Phys. Rev.}
\textbf{\bibinfo{volume}{D84}},
\bibinfo{pages}{011102(R)}
(\bibinfo{year}{2011}).

\bibitem[{\citenamefont{ahmed2}(2011)}]{cdms2011}
\bibinfo{author}{\bibfnamefont{Z.} \bibnamefont{Ahmed}}, \bibnamefont{et al.}:
\bibinfo{journal}{Phys. Rev. Lett.}
\textbf{\bibinfo{volume}{106}},
\bibinfo{pages}{131302}
(\bibinfo{year}{2011}).

\bibitem[{\citenamefont{agnese}(2013)}]{cdms2013}
\bibinfo{author}{\bibfnamefont{R.} \bibnamefont{Agnese}}, \bibnamefont{et al.}:
\bibinfo{journal}{arXiv:1304.4279 [hep-ex]}
(\bibinfo{year}{2013}).

\bibitem[{\citenamefont{Armengaud}(2011)}]{edel2011}
\bibinfo{author}{\bibfnamefont{E.} \bibnamefont{Armengaud}}, \bibnamefont{et al.}:
\bibinfo{journal}{Phys. Lett.}
\textbf{\bibinfo{volume}{B702}},
\bibinfo{pages}{329}
(\bibinfo{year}{2011}).

\bibitem[{\citenamefont{Armengaud}(2011)}]{edelite}
\bibinfo{author}{\bibfnamefont{E.} \bibnamefont{Armengaud}}, \bibnamefont{et al.}:
\bibinfo{journal}{Phys. Lett.}
\textbf{\bibinfo{volume}{B702}},
\bibinfo{pages}{329}
(\bibinfo{year}{2011}).

\bibitem[{\citenamefont{Akimov}(2012)}]{zeplin}
\bibinfo{author}{\bibfnamefont{D.Yu.} \bibnamefont{Akimov}}, \bibnamefont{et al.}:
\bibinfo{journal}{Phys. Lett.}
\textbf{\bibinfo{volume}{B709}},
\bibinfo{pages}{14}
(\bibinfo{year}{2012}).

\bibitem[{\citenamefont{Angle}(2008)}]{xenon10}
\bibinfo{author}{\bibfnamefont{J.} \bibnamefont{Angle}}, \bibnamefont{et al.}:
\bibinfo{journal}{Phys. Rev. Lett.}
\textbf{\bibinfo{volume}{100}},
\bibinfo{pages}{021303}
(\bibinfo{year}{2008}).

\bibitem[{\citenamefont{Akerib}(2013)}]{lux}
\bibinfo{author}{\bibfnamefont{D.S.} \bibnamefont{Akerib}}, \bibnamefont{et al.}:
\bibinfo{journal}{arXiv:1310.8214 [astro-ph.CO]}
(\bibinfo{year}{2013}).

\bibitem[{\citenamefont{McCabe}(2010)}]{mccabe}
\bibinfo{author}{\bibfnamefont{C.} \bibnamefont{McCabe}}:
\bibinfo{journal}{Phys. Rev.}
\textbf{\bibinfo{volume}{D82}},
\bibinfo{pages}{023530}
(\bibinfo{year}{2010}).

\bibitem[{\citenamefont{Green}(2012)}]{green2}
\bibinfo{author}{\bibfnamefont{A.M.} \bibnamefont{Green}}:
\bibinfo{journal}{Mod. Phys. Lett.}
\textbf{\bibinfo{volume}{A27}},
\bibinfo{pages}{12300004}
(\bibinfo{year}{2012}).

\bibitem[{\citenamefont{gondolo}(2012)}]{gondolo}
\bibinfo{author}{\bibfnamefont{P.}~\bibnamefont{Gondolo}},
\bibinfo{author}{\bibfnamefont{G.B.} \bibnamefont{Gelmini}}:
\bibinfo{journal}{JCAP}
\textbf{\bibinfo{volume}{1212}},
\bibinfo{pages}{015}
(\bibinfo{year}{2012}).

\bibitem[{\citenamefont{Iocco}(2010)}]{iocco}
\bibinfo{author}{\bibfnamefont{F.}~\bibnamefont{Iocco}}, \bibnamefont{et al.}:
\bibinfo{journal}{JCAP}
\textbf{\bibinfo{volume}{11}},
\bibinfo{pages}{029}(\bibinfo{year}{2011}).

\bibitem[{\citenamefont{Salucci}(2010)}]{salucci}
\bibinfo{author}{\bibfnamefont{P.}~\bibnamefont{Salucci}}, \bibnamefont{et al.}:
\bibinfo{journal}{Astron. Astrophys.}
\textbf{\bibinfo{volume}{523}},
\bibinfo{pages}{A83}(\bibinfo{year}{2010}).

\bibitem[{\citenamefont{Catena}(2009)}]{catena}
\bibinfo{author}{\bibfnamefont{R.}~\bibnamefont{Catena}},
\bibinfo{author}{\bibfnamefont{P.} \bibnamefont{Ullio}}:
\bibinfo{journal}{JCAP} \textbf{\bibinfo{volume}{08}},
\bibinfo{pages}{004} (\bibinfo{year}{2010}).

\bibitem[{\citenamefont{Freese}(2012)}]{freese}
\bibinfo{author}{\bibfnamefont{K.}~\bibnamefont{Freese}}, \bibnamefont{et al.}:
\bibinfo{journal}{arXiv:1209.3339 [astro-ph.IM]}
(\bibinfo{year}{2012}).

\bibitem[{\citenamefont{Weiner}(2011)}]{weiner}
\bibinfo{author}{\bibfnamefont{P.J.}~\bibnamefont{Fox}},
\bibinfo{author}{\bibfnamefont{J.} \bibnamefont{Liu}},
\bibinfo{author}{\bibfnamefont{N.} \bibnamefont{Weiner}}:
\bibinfo{journal}{Phys. Rev.}
\textbf{\bibinfo{volume}{D83}},
\bibinfo{pages}{103514}
(\bibinfo{year}{2011}).

\bibitem[{\citenamefont{frandsen}(2012)}]{frandsen}
\bibinfo{author}{\bibfnamefont{M.T.}~\bibnamefont{Frandsen}}, \bibnamefont{et al.}:
\bibinfo{journal}{JCAP}
\textbf{\bibinfo{volume}{1201}},
\bibinfo{pages}{024}
(\bibinfo{year}{2012}).

\end{thebibliography}

\end{document}